\documentclass[pdflatex,sn-mathphys-ay,referee]{sn-jnl}

\usepackage{graphicx}%
\usepackage{multirow}%
\usepackage{amsmath,amssymb,amsfonts}%
\usepackage{amsthm}%
\usepackage{mathrsfs}%
\usepackage[title]{appendix}%
\usepackage{xcolor}%
\usepackage{textcomp}%
\usepackage{manyfoot}%
\usepackage{booktabs}%
\usepackage{algorithm}%
\usepackage{algorithmicx}%
\usepackage{algpseudocode}%
\usepackage{listings}%

\usepackage{tabularx}
\usepackage{csquotes}
\usepackage{ragged2e}
\usepackage[mono=true]{libertinus}
\usepackage{eulervm}
\usepackage[scaled=0.9]{zi4}

\let\orcidlogo\undefined
\usepackage{orcidlink}

\newcolumntype{L}{>{\RaggedRight\arraybackslash}X}

\theoremstyle{thmstyleone}%

\theoremstyle{thmstyletwo}%

\theoremstyle{thmstylethree}%

\begin{document}

\title[Multiverse Analysis for CSS]{Making Uncertainty Visible: Multiverse Analysis for Robust Computational Social Science}

\author*[1]{\fnm{Maximilian} \sur{Linde}\orcidlink{0000-0001-8421-090X}}\email{maximilian.linde@gesis.org}

\author[1]{\fnm{Jun} \sur{Sun}\orcidlink{0000-0002-4789-7316}}\email{jun.sun@gesis.org}

\author[1]{\fnm{Paul} \sur{Balluff}\orcidlink{0000-0001-9548-3225}}\email{paul.balluff@gesis.org}

\author[1]{\fnm{Danica} \sur{Radovanovi{\'{c}}}\orcidlink{0000-0003-4245-9031}}\email{danica.radovanovic@gesis.org}

\author[1]{\fnm{Chung-hong} \sur{Chan}\orcidlink{0000-0002-6232-7530}}\email{chung-hong.chan@gesis.org}

\affil[1]{\orgdiv{Department Computational Social Science}, \orgname{GESIS - Leibniz Institute for the Social Sciences}, \orgaddress{\street{Unter Sachsenhausen 6-8}, \city{Cologne}, \postcode{50667}, \country{Germany}}}

\abstract{Through case studies, we demonstrate how multiverse analysis can strengthen the robustness and transparency of computational social science findings against alternative methodological decisions. We conduct multiverse analyses of three published social science studies that use the following computational methods: Bayesian analysis, network generative modeling, and machine learning with or without large language models. These methods are applied frequently in computational social science studies, yet entail a greater degree of arbitrariness in terms of methodological choices, or ``researcher degrees of freedom.'' Our multiverse analyses reveal how the empirical findings in these studies vary as a function of various plausible decision combinations. Our three case studies also expose an often-ignored motivation for conducting multiverse analysis: Showing which methodological combinations lead to computational failure. These failed cases are usually not communicated in the published reports, even though these sophisticated computational methods have a much higher likelihood of failure. We end our paper with suggestions on how to find defensible decision combinations for multiverse analysis of computational social science studies and how to communicate multiverse analysis findings fairly.}


\maketitle

\section{Introduction}\label{introduction}

Computational social science (CSS), as defined by \cite{EdelmannWolffMontagne_2020}, is ``an interdisciplinary field that advances theories of human behavior by applying computational techniques to large datasets from social media sites, the Internet, or other digitized archives such as administrative records.'' (p.~62) Advances in analyzing data with computational methods, such as machine learning (ML) for textual data and generative modeling for network data, have enabled researchers to address research questions that were previously difficult to investigate. However, compared to other quantitative approaches to social science, such as experiments and surveys, tools and workflows used in CSS introduce additional complexity and therefore greater analytical flexibility. For example, big data and digital trace data typically must be preprocessed thoroughly, parameters of advanced modeling strategies must be tuned and set, and even the specific models must be chosen. Consequently, findings reported in CSS papers are likely to be overly dependent on specific decisions that were made during data collection, data preprocessing, and modeling.

The phenomenon that results and conclusions vary depending on which decisions were made in the entire research process was termed \textit{researcher degrees of freedom} in a seminal paper by \cite{SimmonsNelsonSimonsohn2011}. This has been demonstrated further in many-analysts studies \cite[e.g.,][]{SilberzahnUhlmannMartin_2018, BreznauRinkeWuttke_2022}, where different researchers are given a dataset and a research question and independently conduct data preprocessing, statistical analyses, and interpretation of results. These many-analysts studies demonstrate that researchers follow different strategies to solve these tasks and obtain different results and answers to the research question. \cite{SimmonsNelsonSimonsohn2011} suggest six rules that authors must follow to remedy the unacceptably high rate of variation, which mostly focus on transparency. For example, they propose that researchers must disclose a data collection plan, all collected variables, all experimental conditions, and data filtering steps. A similar line of argumentation comes from advocates of preregistration, registered reports, and related instruments \citep{Chambers2013, NosekBeckCampbell_2019, NosekEbersoleDehaven_2018a, LakensMesquidaRasti_2024}, where the idea is that researchers must disclose their decisions prior to collecting and analyzing the data.

We strongly endorse these measures of transparency, preregistration, registered reports, and open science in general. Nevertheless, they still concentrate on only one combination of decisions within the research process, and fail to acknowledge the value of investigating how results would vary for different decision combinations. Multiverse analysis \citep[e.g.][]{SteegenTuerlinckxGelman_2016} addresses this shortcoming by considering all plausible decision combinations in a given research project \citep[see also][as well as similar methods and their definitions in Table \ref{tab:definitions}]{DelgiudiceGangestad2021, PipalSongBoomgaarden2023,short:2025:M,young2025multiverse}. More specifically, the idea behind multiverse analysis is to (1) list all decisions that could possibly be varied, (2) find legitimate alternatives for these decisions, (3) extract plausible decision combinations and filter out implausible or invalid ones\footnote{If it is not feasible to consider all plausible decision combinations, a desired number of decision combinations can be drawn through random sampling \citep{simonsohn:2020:S}.}, and (4) conduct analyses for each remaining decision combination. The findings provide a more complete picture of the research question at hand. They not only provide more transparency but also help to establish or question the robustness of research claims against alternative methodological decisions \citep{athey:2015:MRM}. In addition, multiverse analysis often provides valuable insights into boundary conditions, illuminating the circumstances under which a given finding holds.

Similar methods, listed in Table \ref{tab:definitions}, have a common goal to systematically examine how alternative but defensible analytical choices shape empirical conclusions. However, they differ in which part of the research pipeline they focus on and how the output is presented and understood. In this paper, we follow \cite{short:2025:M} and \cite{young2025multiverse} to subsume all these methods under the arguably most popular label of \textit{multiverse analysis}.

\begin{table}[!ht]
\centering
\caption{List of similar methods and their definitions.} 
\label{tab:definitions}
\begin{tabularx}{\textwidth}{p{4cm} X p{0.5cm}}
\toprule
Method & Definition \\ \midrule
    Multiverse analysis \newline\citep[][p.~703]{SteegenTuerlinckxGelman_2016} & \enquote{Performing the analysis of interest across the whole set of data sets that arise from different reasonable choices for data processing} \newline\\
    Multimodel analysis\newline \citep[][p.~3]{young:2016:MUR} & \enquote{Estimat[ing] the modeling distribution of estimates across all combinations of possible controls as well as specified functional form issues, variable definitions, standard error calculations, and estimation command}\newline\\
    Multiverse-of-methods analysis \newline\citep[][p.~1158]{harder:2020:MM} & \enquote{[Constructing] multiverse of data sets, which is composed of real data sets from studies varying in data-collection methods of interest}\newline\\
    Specification curve analysis \newline\citep[][p.~1208]{simonsohn:2020:S} & \enquote{(1) Identifying the set of theoretically justified, statistically valid and non-redundant specifications; (2) displaying the results graphically, allowing readers to identify consequential specifications decisions; and (3) conducting joint inference across all specifications}\newline\\
    Vibration analysis \newline\citep[][p.~268]{klau:2020:E} & \enquote{Quantify[ing] the variability of results through a vibration ratio, which we defined as the ratio of the largest vs the smallest effect estimate for the same association of interest under different analysis choices}\newline\\
    Computational model robustness analysis \newline\citep[][p.~7]{munoz:2018:WRB} & \enquote{Test[ing] the stability of an estimate across all unique combinations of plausible model ingredients within a theoretically informed model space}\newline\\
    Multiversal method \newline\citep[][p.~1454]{cantone:2023:T} & \enquote{Involving (1) a phase of collection of a systematically differentiated multiplicity of alternative specifications of the same unitary regression model, and (2) a comparative evaluation of fit statistics regarding the estimation of the proprieties of the regression coefficient of the main regressor variable, commonly referred as 'effect size' in studies oriented towards causal inference}\newline\\
    Extreme bounds analysis \newline\citep{leamer1983let} & \enquote{Summarizing the uncertainty about coefficient values in a linear regression model arising from uncertainty about the correct set of regressors in the equation}\newline\\
\bottomrule
\end{tabularx}
\vspace{5pt} 
    \footnotesize
    \textit{Note:} Extreme bounds analysis was first proposed in \cite{leamer1983let}. The definition was from \cite[][p.~182]{magee1990asymptotic}.
\end{table}

\subsection{Does CSS Need Multiverse Analysis?}

Articles on multiverse analysis typically motivate its use with examples based on regression analysis of observational data. For instance, the controversial study by \cite{jung:2014:F}, which reported that hurricanes with feminine names caused more damage than those with masculine names, was later shown to be non-robust through a multiverse analysis \citep{simonsohn:2020:S}, in which alternative specifications for inclusion criteria, operationalizations of dependent and independent variables, and specifications of regression models were explored. \cite{young2025multiverse} argue that the analysis of observational data naturally encompasses more researcher degrees of freedom than the analysis of experimental data for two main reasons: first, methodological abundance (multiple defensible ways exist to model the same data); and second, model uncertainty (models may capture methodological artifacts rather than underlying reality). Building on this argument, we contend that CSS research exhibits an even higher level of researcher degrees of freedom than other quantitative approaches in social science, and especially than other non-computational areas of science.

In a prototypical CSS study, researchers often rely on digital trace data (e.g., from social media platforms) or online news articles, which are by nature observational.\footnote{Experiments do exist in this domain \citep[e.g.,][]{bond:2012,kramer:2014:E}. However, such large-scale experiments are rarely conducted by academic researchers and often raise substantial ethical concerns \citep{2014PNASconcerns}.} The use of these data further amplifies the problems of methodological abundance and model uncertainty \citep[cf.][]{young2025multiverse}. For instance, textual data can be analyzed using a wide range of computational methods, including dictionary-based approaches, ML methods such as support vector machines (SVMs), pre-trained transformer models, and more recently large language models (LLMs). On top of that, each of these approaches also entails numerous parameter choices (e.g., learning rates or the selection of pre-trained models). Within the realm of the relatively simple dictionary-based approaches, \cite{chan_4b} identified $37$ different dictionaries to quantify sentiment in news articles and demonstrated that switching dictionaries can substantially alter both the magnitude and the direction of empirical findings. More advanced computational approaches do not resolve the problem. For example, \cite{baumann:2025:LLM} similarly showed that researchers can obtain different conclusions by switching between LLMs or prompting strategies. Because many of these decisions are both arbitrary and defensible, multiverse analysis is essential for demonstrating how such choices influence empirical findings.

Furthermore, model uncertainty in CSS differs fundamentally from the uncertainty arising from model misspecification in regression analysis. In many ML-based CSS studies, researchers employ a two-step workflow, in which ML models are used primarily for creating theoretically relevant variables based on unstructured data such as texts; and the resulting variables are subsequently analyzed using statistical methods such as regression modeling. This two-step workflow introduces a unique form of model uncertainty. Specifically, validation of ML prioritizes discriminative metrics (e.g., $F_1$ scores) while often overlooking model calibration (e.g., Brier scores). Such neglect is consequential because misclassifications are seldom random; instead, they often correlate with both observed and unobserved variables embedded in complex causal structures (e.g., confounders, moderators, mediators, colliders), thereby inducing systematic measurement error in second-step regression results \citep{fong:2020:MLP,teblunthuis:2024:MAC}.\footnote{This problem is not unique to ML-based studies; it also exists in many study designs, for instance linkage study using content analytic methods linked with survey, which looks for ecological associations \citep[e.g.,][]{scharkow:2016:HME}. Only ML makes this kind of multistep design more common and the measurement problem more salient.} Given that fully specifying these causal structures is impossible \citep{linde:2025:RSU}, multiverse analysis becomes essential for revealing how different ML procedures and their associated error patterns propagate to the following steps and influence empirical outcomes.

Another argument for multiverse analysis in CSS is the identification of decision combinations that result in computational issues. Many computational methods used in CSS, such as transformers, exponential random graph modeling (ERGM), and Bayesian multilevel modeling, are algorithmically complex and may suffer from convergence problems or fail entirely for certain combinations of data and model parameters. These computational issues are often difficult to diagnose and resolve. In practice, CSS researchers may switch algorithms, settings, priors, or tuning parameters to circumvent these computational issues.\footnote{Researchers are obviously well-advised to investigate the sources of the computational issues. Very often, these issues can be informative about model misspecifications.} Although these decisions are often practical rather than arbitrary, they are rarely documented transparently in published manuscripts. A multiverse analysis can reveal which decision combinations lead to computational issues or failures, which can then be reported explicitly rather than being omitted.

In summary, CSS strongly benefits from multiverse analysis as a tool to foster transparency. The field is characterized by big and complex data, an abundance of analytical methods, substantial model uncertainty, and an elevated risk of modeling failure due to computational constraints.

\subsection{Why Is Multiverse Analysis Rarely Practiced in CSS?}

Even though multiverse analysis has the potential to enrich studies by offering a more transparent and comprehensive account of decision combinations, it remains rarely implemented in CSS \citep[cf.][]{PipalSongBoomgaarden2023,ivanusch:2025:messages}. More broadly, multiverse analysis is still a marginal approach within the social sciences \citep{rijnhart:2021:ARM}, and its application in computational settings generates additional difficulties.

At its core, multiverse analysis involves repeating similar analyses across a range of different decision combinations. Even when considering only one decision combination, which is conventionally done, CSS research is often computationally demanding. The adoption of newer methods such as LLMs further intensifies this demand, as they typically require either special hardware or paid access to commercial services. These constraints entail substantial costs in terms of time, financial resources, and environmental impact \citep[e.g.,][]{JeghamAbdelattiKoh_2025}. Because multiverse analysis essentially multiplies the number of analyses, it also multiplies these resource requirements, making it unappealing for researchers who do not have excess computational resources. This reluctance to use multiverse analyses might even be amplified by the erroneous impression that writing the code for a multiverse analysis is complex and time-consuming.

In this article, we advocate an eclectic position to balance the need for multiverse analysis in CSS (previous section) and the practical difficulties of conducting multiverse analysis in CSS (this section). Instead of advocating to include a broad range of defensible decision combinations, we argue that researchers should focus on a selective subset of the most important paths. In neuroscience, this approach is called ``mini-multiverse analysis'' or ``mini-verse analysis'' \citep{mullin:2025:R}.\footnote{All research, not only within CSS and neuroscience, entails a potentially vast, if not infinite, set of defensible decision combinations. For instance, there are innumerable ways to draw a random sample from a dataset. Only computational methods in CSS make this fact more visible. Accordingly, all multiverse analyses necessarily involve selection and could, in this sense, be understood as ``mini-verse analyses.''} In line with prior suggestions by \cite{PipalSongBoomgaarden2023}, preregistering these important decision combinations is essential. This is because, ironically, the selection of ``important'' decision combinations in a multiverse analysis itself constitutes an additional layer of researcher degrees of freedom.

\subsection{The Present Paper}

The present paper demonstrates, through three case studies, how multiverse analysis can be used in typical CSS research for evaluating the robustness of empirical outcomes in a transparent and principled manner. The three case studies are: Bayesian analysis in \cite{ChanRauchfleisch2023}, network generative modeling in \cite{beffel2023understanding}, and ML in \cite{Birkenmaier2025}. These three studies were chosen because they reflect a diverse set of research methods in CSS, because all materials (i.e., data and code) were publicly available, and because they were recently published.

\section{Case Study 1: \cite{ChanRauchfleisch2023}}\label{case_study_1}

\subsection{The Original Study}

As an example to demonstrate the utility of Bayesian analysis for journalism studies, \cite{ChanRauchfleisch2023} investigated the amount of China news coverage in countries as a function of their distance to China \citep[this refers to section ``Example~2: useNews'' in][]{ChanRauchfleisch2023}. For their main analysis, the authors operationalized distance as the log trading volume between countries and China. Their (preregistered) hypothesis was that there is a positive effect of distance, in terms of log trading volume, on China news coverage. For their analysis, they used the useNews dataset \citep{PuschmannHaim2020}. Specifically, they used the $2019$ media content data that contains $76$ media outlets around the world, collected from MediaCloud. They removed all outlets that have less than $1,000$ articles, yielding a total of $61$ outlets for further consideration. 

Subsequently, they used a dictionary approach to check whether one of the keywords ``chinese'', ``china'', ``beijing'', or ``shanghai'' is present within the $1,525,871$ news articles from the $61$ outlets. If at least one match was recorded for a given article, this was counted as coverage of China. That way, for each outlet, a count of articles covering China is obtained out of all articles.

Trading data for China in $2019$ were obtained from the China statistical yearbook 2019 \citep{Nationalbureauofstatisticschina2019}. The data contains, among other things, information about the import and export trading volume of countries with respect to China. As such, since each outlet is situated in some country, information about trading volumes of that country with China is obtained.

To estimate the relationship between import trading volume and China news coverage, \cite{ChanRauchfleisch2023} chose to use a Bayesian multilevel regression model. They modeled China news coverage with a negative binomial likelihood and a log link. Moreover, since the $61$ outlets are nested within $11$ countries, they included random intercepts for countries. Standard Gaussian priors, $\mathcal{N}\left(0, 1\right)$, were used for both the grand intercept and the fixed effect of log import, whereas the other parameters followed the default priors determined by the R package brms \citep{Bxrkner2017, Bxrkner2018}.

The results of their Bayesian multilevel regression suggested a positive relationship between trading volume and China news coverage. They reported a coefficient of $0.25$ with a $95\%$ credible interval (CrI) of $\left[0.15, 0.34\right]$. That is, increasing log import trading volume by $1$ increases the log rate of China news coverage by $0.25$. These findings support the authors' hypothesis.

\subsection{Multiverse Analysis}

The study by \cite{ChanRauchfleisch2023} contains countless decisions, which become evident when studying the details of their work. Some of these decisions are well-explained, such as the decision to log-transform the import trading volume because it has been shown that the relationship between news coverage and trading volume is not linear \citep[cf.][]{Grasland2020, Wu2000}. Other decisions were potentially made more implicitly, such as the use of default priors for the variation of the random intercepts and the shape parameter. Independent of whether decisions have been made implicitly or explicitly, it can be claimed that there are several plausible alternatives to some decisions. We used multiverse analysis to investigate the robustness of the results of \cite{ChanRauchfleisch2023} as a function of using alternative decision combinations.

In our multiverse analysis of \cite{ChanRauchfleisch2023}, we considered six decisions for which we thought that it is plausible that they could be made differently: (1) the specification of the Chinese dictionary used for matching articles; (2) the minimum number of Chinese dictionary matches that must be present to categorize a given article to cover China; (3) the minimum number of articles in an outlet to consider the outlet at all; (4) which type of predictor to use as a measure of distance between countries and China; (5) the statistical model used to estimate China news coverage; and (6) the prior for the fixed effect in the Bayesian hierarchical model. These six decisions, together with the original choice in \cite{ChanRauchfleisch2023} and our variations thereof, yielded $144$ decision combinations and can be found in Table~\ref{tab:chan_multiverse_decisions}.
\begin{table}[!ht]
\centering
\caption{Overview of the decision combinations for the multiverse analysis of~\cite{ChanRauchfleisch2023}: analytical decision considered and their defensible options.} 
\label{tab:chan_multiverse_decisions}
\begin{tabularx}{\textwidth}{p{6cm} X p{0.5cm}}
\toprule
Analytical decision & Defensible options \\ \midrule
Dictionary & (1) [chinese, china, beijing, shanghai]\newline
             (2) [chinese, china]\\
Minimum number of dictionary matches & (1) $1$ \newline
                                       (2) $10$\\
Minimum number of articles & (1) $1000$\newline
                             (2) $0$ \\
Predictor & (1) $\log\left(\text{import}\right)$\newline
            (2) $\log\left(\text{export}\right)$\newline
            (3) $\left(\log(\text{import}) + \log(\text{export})\right) / 2$\\
Statistical model & (1) Negative binomial \\
& (2) Poisson\\
Prior & (1) $\mathcal{N}(0, 1)$\newline
        (2) $\mathcal{N}(0, 0.5)$\newline
        (3) $\mathcal{N}(0, 2)$\\
\midrule
Total specifications & $2 \times 2 \times 2 \times 3 \times 2 \times 3 = 144$ \\ \bottomrule
\end{tabularx}
\vspace{5pt} 
    \footnotesize
    \textit{Note:} Defensible option (1) for each analytical decision is the original setting.
\end{table}

\subsection{Results}

\begin{figure}[!ht]
    \centering
    \includegraphics[width=\linewidth]{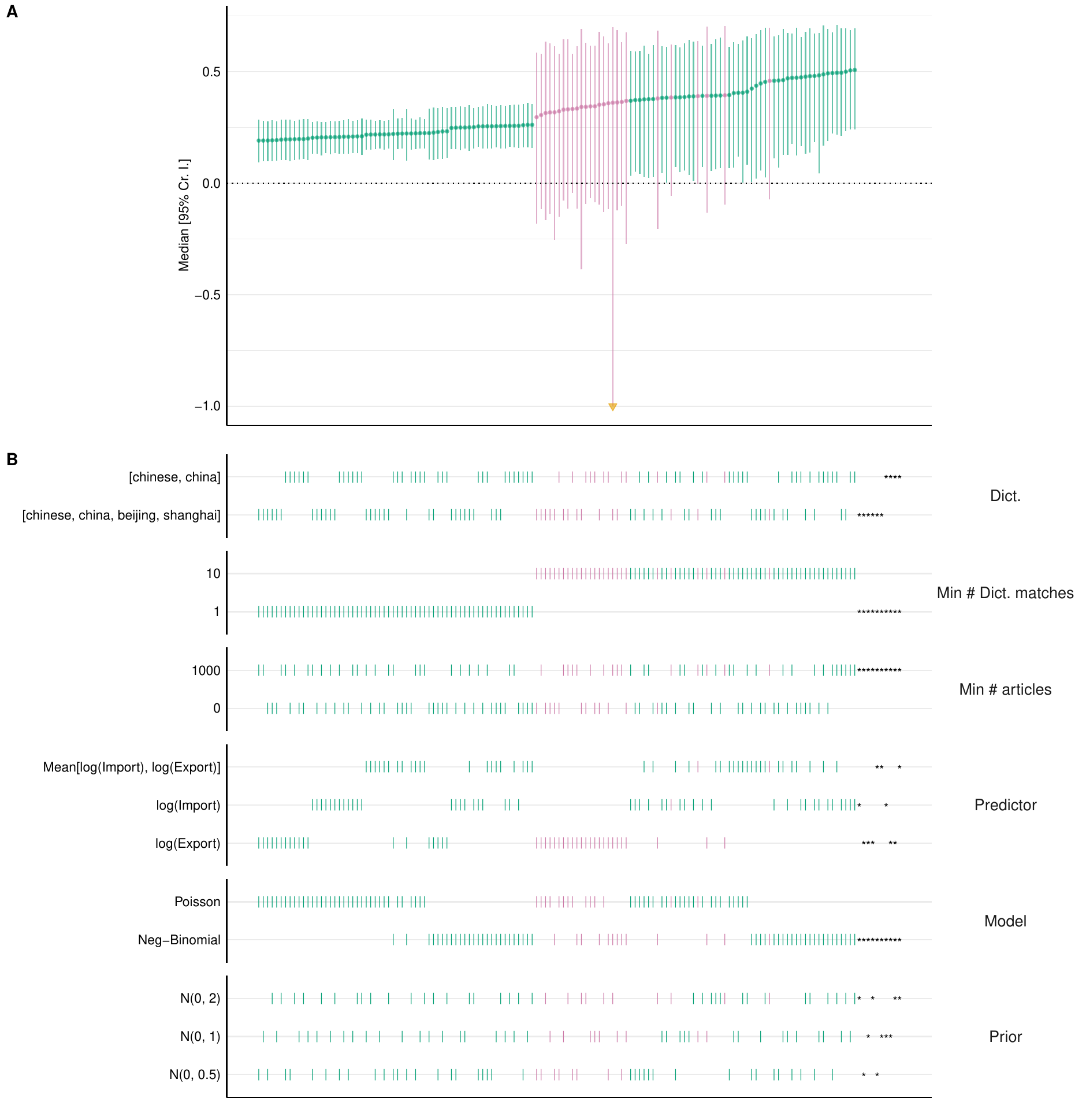}
    \caption{Results of the multiverse analysis of \cite{ChanRauchfleisch2023}. Panel~A displays posterior median estimates and corresponding $95\%$ CrIs for the fixed effect as vertical bars. Panel~B shows the specific decision combinations that were used for a specific effect shown in Panel~A. Green represents decision combinations for which the $95\%$ CrI is positive (same as the original direction); pink represents decision combinations for which the $95\%$ CrI overlaps with $0$; the orange arrows in Panel~A indicate that the lower or upper limit of the $95\%$ CrI are truncated for purposes of visualization; the black stars in Panel~B show decision combinations for which the Bayesian models could not be fit due to computational issues.}
    \label{fig:chan_multiverse}
\end{figure}

The results of our multiverse analysis of \cite{ChanRauchfleisch2023} are shown in Figure~\ref{fig:chan_multiverse}. It is evident that all of the posterior median estimates of the fixed effect are positive, going in the same direction as in the original study. Most of the associated $95\%$ CrIs do not overlap with $0$ (green) but there are also some that do (pink). Panel~B suggests that pink findings are almost exclusively found when log export was used as a predictor and when the minimum number of Chinese dictionary matches was $10$. Furthermore, the black stars on the right side of Panel~B demarcate models that encountered computational issues and could therefore not be fit. Those belong exclusively to decision combinations in which a negative-binomial model was used, coupled with a minimum number of Chinese dictionary matches of $1$ and a minimum number of articles of $1000$. Overall, the results of our multiverse analysis suggest that the findings of \cite{ChanRauchfleisch2023} are robust. None of the estimated effects were in the opposite direction compared to the original findings. If anything, our results even suggest larger effects than the original study, since around $66\%$ of our results yielded posterior median estimates larger than $0.25$.

It is worthwhile to take a closer look at the decision combinations that encountered computational issues (see black stars in Panel~B of Figure~\ref{fig:chan_multiverse}). More precisely, sampling from the posterior was never finished but instead the algorithm got stuck at some iteration.\footnote{Not only did the algorithm for sampling from the posterior get stuck but also R/RStudio froze. We circumvented this problem by implementing a timeout exit routine within subprocesses.} Computational issues like this are sometimes encountered when fitting Bayesian models. In some cases, they might point to some underlying problem with the model specification, which should be solved first. Often, however, it becomes evident after inspecting diagnostic information that the model specification is not flawed.

\begin{figure}[!ht]
    \centering
    \includegraphics[width=\linewidth]{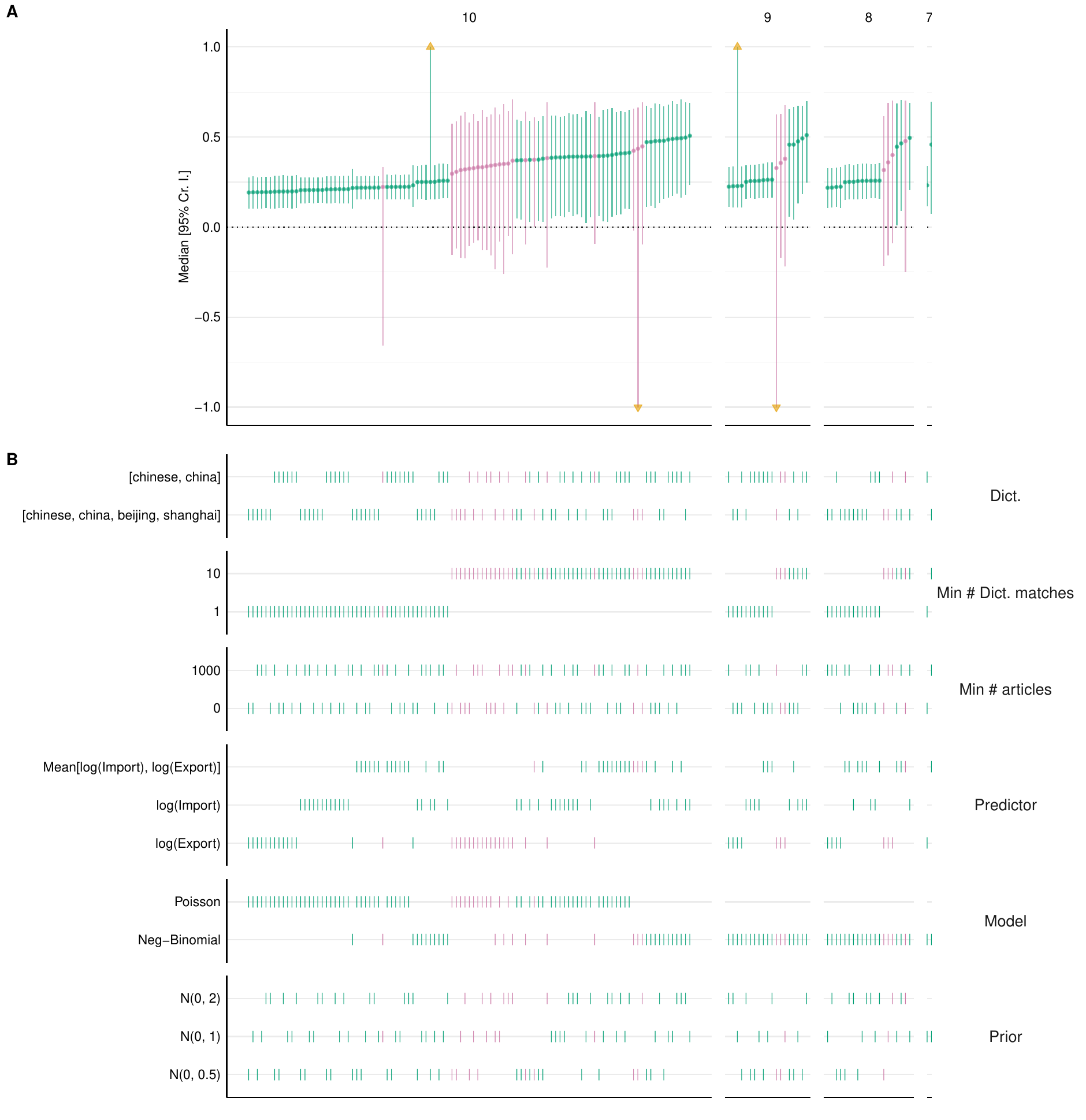}
    \caption{Results of the multiverse analysis of \cite{ChanRauchfleisch2023}. Panel~A displays posterior median estimates and corresponding $95\%$ CrIs for the fixed effect as vertical bars. The values were obtained from the combination of all posterior samples of the successful models out of $10$ trials. The columns display how many of the $10$ trials successfully finished. Panel~B shows the specific decision combinations that were used for a specific effect shown in Panel~A. Green represents decision combinations for which the $95\%$ CrI is positive (same as the original direction); pink represents decision combinations for which the $95\%$ CrI overlaps with $0$; the orange arrows in Panel~A indicate that the lower or upper limit of the $95\%$ CrI are truncated for purposes of visualization.}
    \label{fig:chan_multiverse_extended}
\end{figure}

Since we explored an automated procedure, we were forced to discard these non-converging models entirely, without any further inspection of diagnostic information. Nevertheless, it is possible that the set of decision combinations that did not converge was random. To investigate this possibility, we repeated our multiverse analysis with one important change: Each decision combination was fit $10$ times. By doing this, we could examine how many of the $10$ attempts converge and how many do not. The results are shown in Figure~\ref{fig:chan_multiverse_extended}, where, in contrast to Figure~\ref{fig:chan_multiverse}, the plot is divided into several columns. The columns represent how many of the $10$ trials for fitting the Bayesian model succeeded. The results indicate that at least half of the models converged for each decision combination, providing some reassurance that the model specifications were not completely flawed. Further, it is evident that, without exception, occurrences of non-converging models are found when using a negative-binomial likelihood. In general, the results shown in Figure~\ref{fig:chan_multiverse_extended} mirror those in Figure~\ref{fig:chan_multiverse}, such that all effects are in the same direction as the original effect and around $65\%$ of posterior estimates are even larger in magnitude than the originally reported effect.

It is hardly possible to list all decisions that were made and conceive alternative decisions for them. Our choice of six decisions reflects only a small part of the space with all decisions. For example, our variations of the predictor could be expanded to also include physical distance or some measure of cultural distance. Or, taking it to the extreme, for the minimum number of articles in an outlet, any number could be used, not just $1000$ and $0$ as in our multiverse analysis. Moreover, an entirely new decision could be added. No multiverse analysis can claim to be exhaustive.

Nevertheless, many multiverses are extensive and it is hard to evaluate every specification in that multiverse. \cite{RihaSicchaOulasvirta_2024} propose an iterative filtering approach, when working with Bayesian models. Their framework treats the multiverse as a starting point than can be progressively pruned and refined. Researchers begin with the full multiverse an then iteratively filter out models based on diagnostic criteria, predictive performance, or theoretical coherence. Through this, in each iteration the attention is shifted to better-working models, thereby refining the multiverse and making the complexity of multiverse analysis manageable.

\section{Case Study 2: \cite{beffel2023understanding}}\label{case_study_2}

\subsection{The Original Study}

In the original study by \cite{beffel2023understanding}, the authors conducted a social network analysis using Exponential Random Graph Models \citep[ERGMs,][]{robins2007introduction} with classroom social network data and node attribute data (e.g., prosocial score) to examine the effects of (1) prosocial behavior and (2) homophily of prosocial behavior on children's classroom affiliative relationships.

\subsubsection{Data Collection and Preprocessing}

The authors collected social network data and peer-nomination data from $N=677$ students in 34 second- through fourth-grade classrooms across five Chicago public elementary schools.

The social network data was collected using an approach called cognitive social structures \citep{krackhardt1987cognitive}, which was then adapted by \cite{neal2008kracking} for measuring children's perceptions of affiliative relationships in their classroom. In their study, participants completed a social network survey with separate pages for themselves and each classmate, circling any names of classmates each child ``hangs out with often'' in the classroom.
The collected survey data for each classroom were then aggregated following \cite{neal2008kracking}.
The social network data for each classroom in the end was a simple undirected network, where each node represented a participant and an edge between two nodes represented an affiliative relationship between them.

The authors also calculated a ``prosocial score'' for each participant to measure their prosocial behavior based on peer-nomination surveys including three items from the Child Social Behavior Scale--Peer Report \citep[e.g., ``who helps others join a group'';][]{crick1995relational}. For each item, a participant's received nominations (excluding self-nominations) were divided by the total possible nominations to yield proportions, which are then averaged across the three items and rescaled such that the prosocial scores ranged from 0 to 100.

\subsubsection{ERGM and Meta-analysis}
The authors hypothesized that
\begin{itemize}
    \item (Hypothesis 2.1) higher levels of individual prosocial behavior will be associated with an increased likelihood of an affiliative relationship;
    \item (Hypothesis 2.2) children who have similar levels of prosocial behavior will be more likely to have affiliative relationships than children without similar levels of prosocial behavior.
\end{itemize}

With the collected social network data and the prosocial scores of individuals, the authors fit an ERGM for each classroom using the software implementation by \cite{hunter2008ergm}.
As covariates, they included gender as an individual characteristic, and gender homophily using the uniform homophily term, or \texttt{nodematch} in \cite{hunter2008ergm}'s software implementation, as a dyadic characteristic.
To model transitivity in the networks, they included Geometrically Weighted Edgewise Shared Partner (GWESP) in the models with a decay factor of $\alpha = 0.7$ in all but two classrooms, where it was altered to help with model convergence and model fit ($\alpha = 0.5$ and $\alpha = 0.8$, respectively).
To test the two hypotheses, as predictors, they included prosocial score as an individual characteristic, as well as homophily distance of prosocial behavior using the absolute difference in nodal attribute (\texttt{absdiff}) term as a dyadic characteristic which computes the absolute difference between the prosocial scores of the two incident nodes. Negative estimates of \texttt{absdiff} indicate that greater prosocial behavior homophily corresponds to a higher likelihood of having affiliative relationships.

The authors then used a combination of Akaike information criteria (AIC) and Bayesian information criteria (BIC) values, goodness-of-fit plots and MCMC diagnostics to assess the convergence and the goodness of fit of each model. They varied the model configurations (e.g., values of \texttt{MCMC.burnin}, \texttt{MCMC.interval}, \texttt{MCMC.samplesize}) for individual models to help achieve convergence. The random seed was set to $40$ for all models.
Out of the 34 classroom networks, 12 networks showed acceptable to good model fit and convergence.

The authors conducted random-effects multilevel meta-analyses on the 12 ERGMs for the two hypotheses respectively. 
The results supported both hypotheses.
For Hypothesis 2.1, the pooled effect size of prosocial behavior was $b=0.046$; $95\%~\text{CI}=[0.034, 0.059]$.
For Hypothesis 2.2, the pooled effect size for homophily distance of prosocial behavior was $b=-0.024$; $95\%~\text{CI}=[-0.045, -0.003]$.

\subsection{Multiverse Analysis}

The authors of the original study do not provide the raw data but only the preprocessed data (i.e., the aggregated adjacency matrices of classroom social networks and node-level properties such as gender and scaled prosocial scores). Even though many decisions of the data preprocessing could have been included in our multiverse analysis, we have to focus our multiverse analysis on network model specifications and model fitting. We vary analytical decisions in the ERGM specification and fitting across 5 dimensions, leading to $3 \times 2 \times 3 \times 6 \times 3= 324$ decision combinations. The overview of the decision combinations is presented in Table~\ref{table_overview_beffel}.
These five decisions ensure a balance between decisions that affect model specifications (i.e., handling isolate nodes, distance measure for homophily), estimation and convergence (i.e., handling non-converging models), and stochasticity (i.e., random seeds).

\begin{table}[!ht]
\centering
\caption{Overview of the decision combinations for the multiverse analysis of~\cite{beffel2023understanding}: analytical decision considered and their defensible options.
\label{table_overview_beffel}}
\begin{tabularx}{\textwidth}{p{4.5cm} X}
\toprule
Analytical decision & Defensible options \\
\midrule

Handling isolate nodes 
& (1) Including \texttt{isolates} in the model when present$^a$   \newline
  (2) Removing isolate nodes  \newline
  (3) Not including \texttt{isolates} in the model
 \\
Distance measure for homophily 
& (1) Absolute difference  \newline
  (2) Squared difference
\\
Decay factor of GWESP $\alpha$
& (1) Original value  \newline
  (2) Original value $+0.05$  \newline
  (3) Original value $-0.05$
\\
Handling non-converging models 
& (1) \texttt{main.method=MCMLE}, \texttt{MCMLE.termination=confidence}$^b$  \newline
  (2) \texttt{main.method=MCMLE}, \texttt{MCMLE.termination=Hummel}  \newline
  (3) \texttt{main.method=MCMLE}, \texttt{MCMLE.termination=Hotelling}  \newline
  (4) \texttt{main.method=MCMLE}, \texttt{MCMLE.termination=precision}  \newline
  (5) \texttt{main.method=MCMLE}, \texttt{MCMLE.termination=none}  \newline
  (6) \texttt{main.method=Stochastic-Approximation}
\\
Random seeds 
& (1) 40  \newline
  (2) 400  \newline
  (3) 4000
\\
\midrule
Total specifications & $3 \times 2 \times 3 \times 6 \times 3= 324$ \\
\bottomrule
\end{tabularx}
\vspace{5pt} 
    \footnotesize
    \textit{Note:} Defensible option (1) for each analytical decision is the original settings.\newline $^a$ The original authors stated that they ``\textit{included an \texttt{isolates} term in models for which the social network contained isolate nodes}''.
However, we observe that this is not entirely true. For example, class 6013 contains isolate nodes (6305, 6312) but has no \texttt{isolates} term in the model. In our multiverse analysis, we choose to faithfully implement this variation as the original authors stated in the text.\newline
$^b$ The variation used in the original study was implicit. However, it was evident from the default setting in \texttt{ergm} version 4.1.2 which the original study used.
\end{table}

In the original study, 12 out of 34 classroom network models achieved acceptable convergence and goodness-of-fit to the data.
In our multiverse analysis, the number of convergent and well-fit classroom network models varies depending on the decision combination.
Due to the scale of the multiverse analysis (in total $34 \times 324 = 11016$ classroom network models), it is not feasible to rely on manual visual inspection of the MCMC diagnostics plots and goodness-of-fit plots, as was done in the original study. We therefore apply a few rules for assessing model convergence and goodness-of-fit.
First, the \texttt{ergm()} function to fit a model must return with no error and within 120 seconds (in our reproduction of the original study, a model took at most $33.7$ seconds to fit) because an exaggerated fit time is an indicator of model non-convergence.
Second, we exclude models which throw a warning that may indicate bad identifiability, such as non-existing MPLE (Maximum pseudolikelihood estimation), singular approximate Hessian matrix, or linear dependence among the model statistics.
Third, we check the joint Geweke $p$-values in the burn-in diagnostics and filter out model fits with Geweke $p < 0.1$, which indicates bad model convergence.
Fourth, we compare the AIC and BIC values from the null models that contain \texttt{edges} as the only term and the final models. We exclude the final models in which the AIC increases, or in which the BIC increases by more than 10\%.

For each decision combination, we then perform the random-effects multilevel meta-analyses on all remaining ERGMs for the two hypotheses respectively.

\subsection{Results}

\begin{sidewaysfigure}
    \centering
    \includegraphics[width=\linewidth]{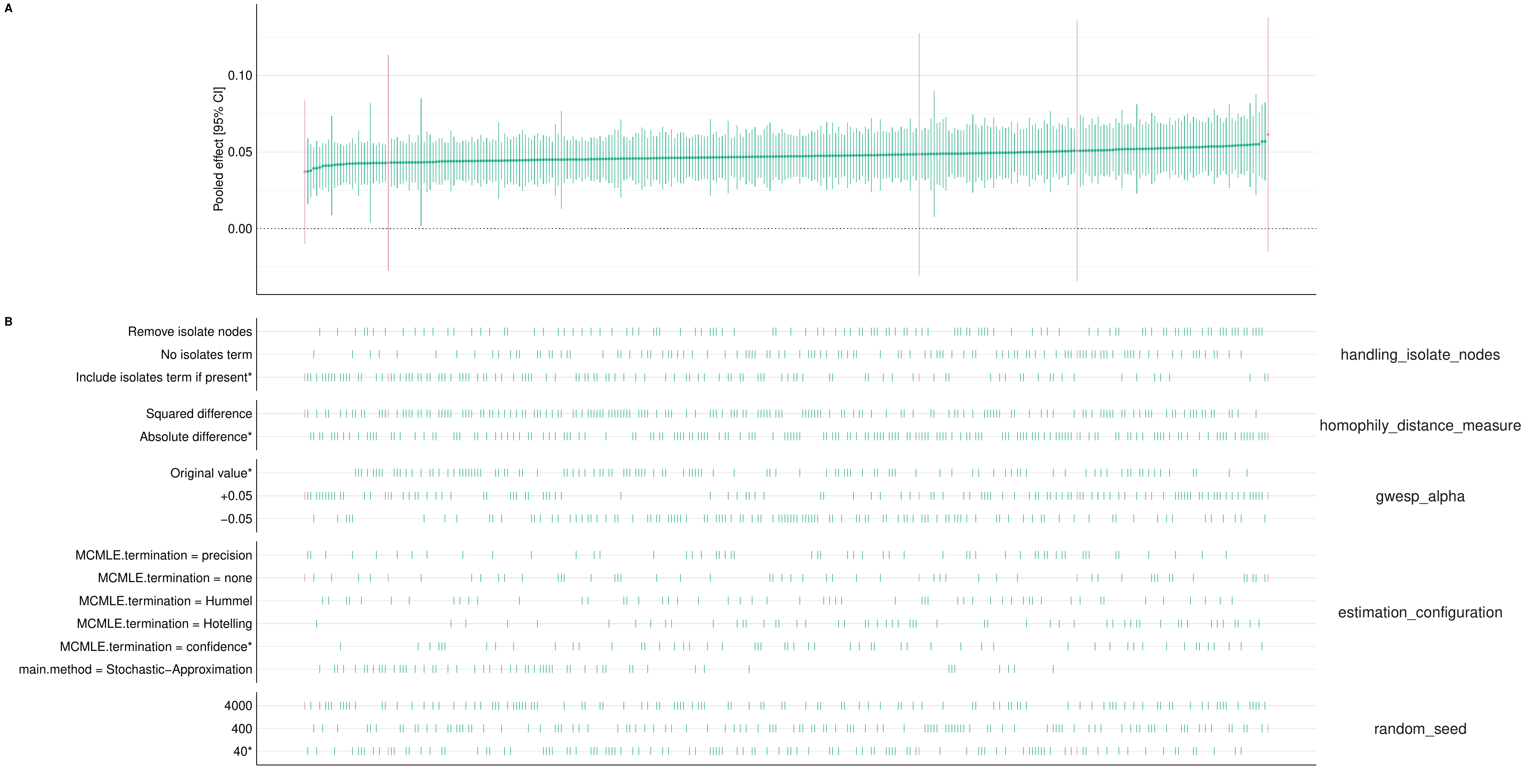}
    \caption{Results of the multiverse analysis of Study 1 in \cite{beffel2023understanding}.
        Panel~A displays the estimates (pooled effect sizes) of the prosocial score from the meta-analysis as dots and the corresponding $95\%$ CIs as vertical bars.
        Panel~B shows the specific decision combinations that were used for a specific effect shown in Panel~A. Green represents decision combinations for which the $95\%$ CI is positive (same as the original direction); pink represents decision combinations for which the $95\%$ CI overlaps with $0$.}
    \label{fig_beffel_multiverse_ps}
\end{sidewaysfigure}

\begin{sidewaysfigure}
    \centering
    \includegraphics[width=\linewidth]{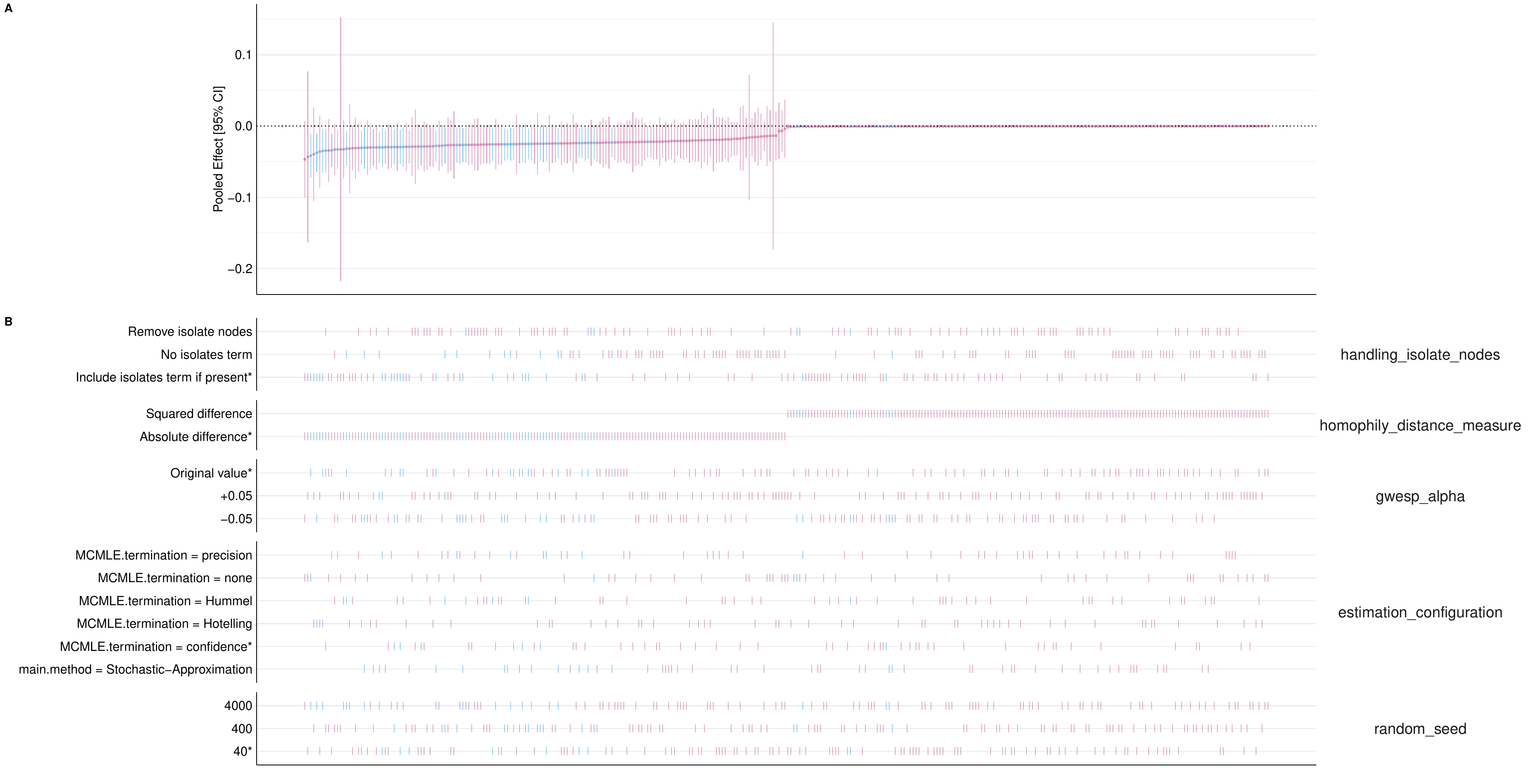}
    \caption{Results of the multiverse analysis of Study 2 in \cite{beffel2023understanding}.
        Panel~A displays the estimates (pooled effect sizes) of the prosocial score homophily distance from the meta-analysis as dots and the corresponding $95\%$ CIs as vertical bars.
        Panel~B shows the specific decision combinations that were used for a specific effect shown in Panel~A. Blue represents decision combinations for which the $95\%$ CI is negative (same as the original direction); pink represents decision combinations for which the $95\%$ CI overlaps with $0$.}
    \label{fig_beffel_multiverse_psh}
\end{sidewaysfigure}

The results of our multiverse analysis of \cite{beffel2023understanding} are shown as specification curves in Figure~\ref{fig_beffel_multiverse_ps} (for Hypothesis 2.1) and Figure~\ref{fig_beffel_multiverse_psh} (for Hypothesis 2.2). 

Figure~\ref{fig_beffel_multiverse_ps} shows that almost all pooled effect size estimates of the prosocial score and their corresponding $95\%$ CIs are positive (green), going in the same direction as in the original study, while only 5 overlap with 0 (pink).
Overall, it is evident that the conclusion about Hypothesis 2.1 in the original study is robust against other decision combinations that are considered in our multiverse analysis.

Figure~\ref{fig_beffel_multiverse_psh} shows the pooled effect size estimates of the prosocial score homophily distance and their corresponding $95\%$ CIs. Only 47 out of 324 $95\%$ CIs are negative (blue), going in the same direction as in the original study, while the remaining overlap with 0.
Panel~B suggests that the most decisive analytical decision is the homophily distance measure.
When we choose to measure homophily distance with squared difference, only 6 out of 162 of the 95\% CIs do not overlap with $0$. When we choose to measure homophily distance with absolute difference (same as in the original study), 41 out of 162 of the 95\% CIs do not overlap with $0$. In addition, squared differences clearly lead to much smaller estimates.
Overall, these results from our multiverse analysis show that Hypothesis 2.2 (i.e., children who have similar levels of prosocial behavior will be more likely to have affiliative relationships than children without similar levels of prosocial behavior) in the original study is suggestive, being sensitive to the variations of analytical decisions that we consider, especially how the similarity levels of prosocial behavior are measured.

\begin{sidewaysfigure}
    \centering
    \includegraphics[width=\linewidth]{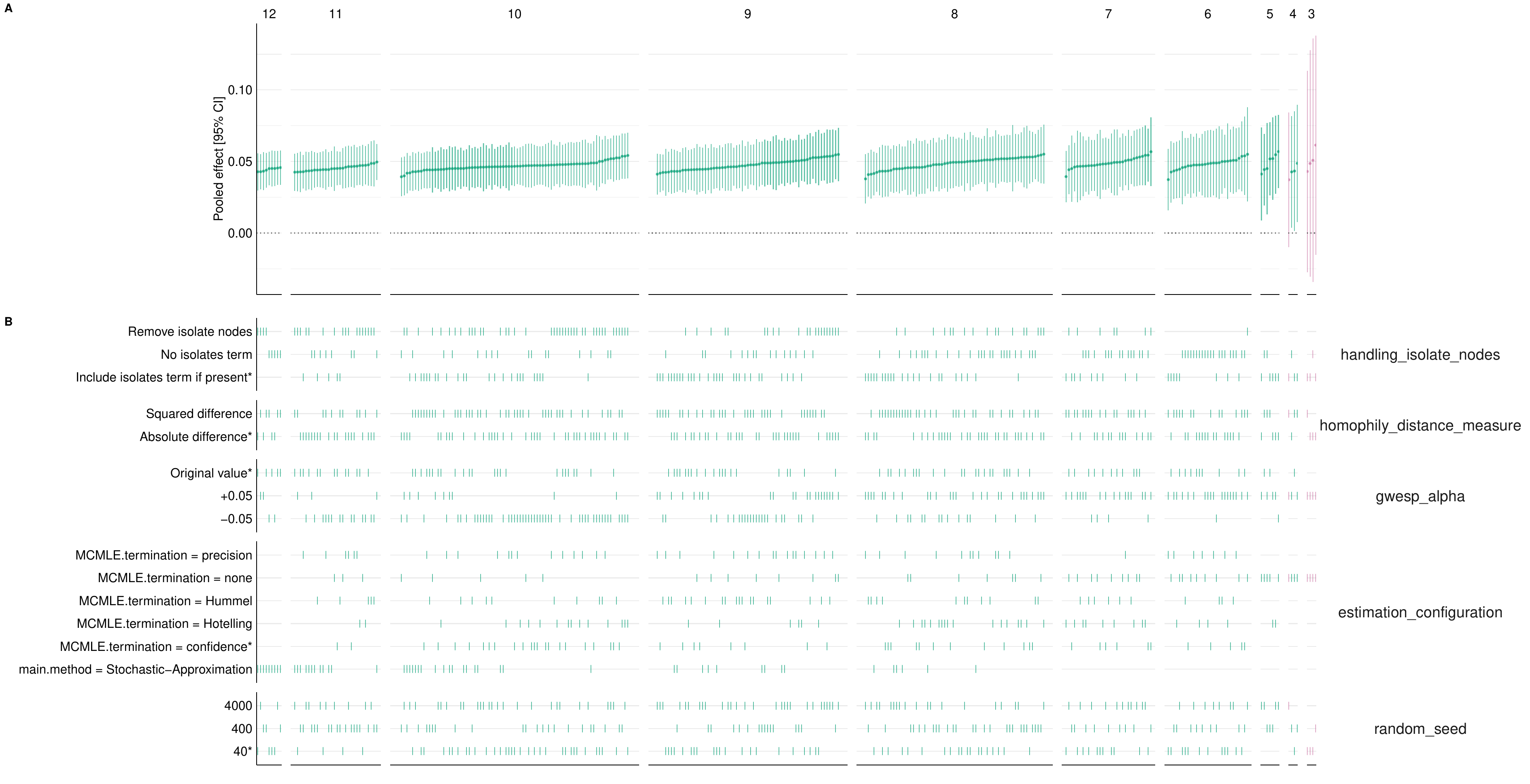}
    \caption{Results of the multiverse analysis of Study 1 in \cite{beffel2023understanding}.
        Panel~A displays the estimates (pooled effect sizes) of the prosocial score from the meta-analysis as dots and the corresponding $95\%$ CIs as vertical bars.
        The columns display how many ERGMs of the 34 classroom networks achieved acceptable convergence and goodness-of-fit.
        Panel~B shows the specific decision combinations that were used for a specific effect shown in Panel~A. Green represents decision combinations for which the $95\%$ CI is positive (same as the original direction); pink represents decision combinations for which the $95\%$ CI overlaps with $0$.}
    \label{fig_beffel_multiverse_extended_ps}
\end{sidewaysfigure}

\begin{sidewaysfigure}
    \centering
    \includegraphics[width=\linewidth]{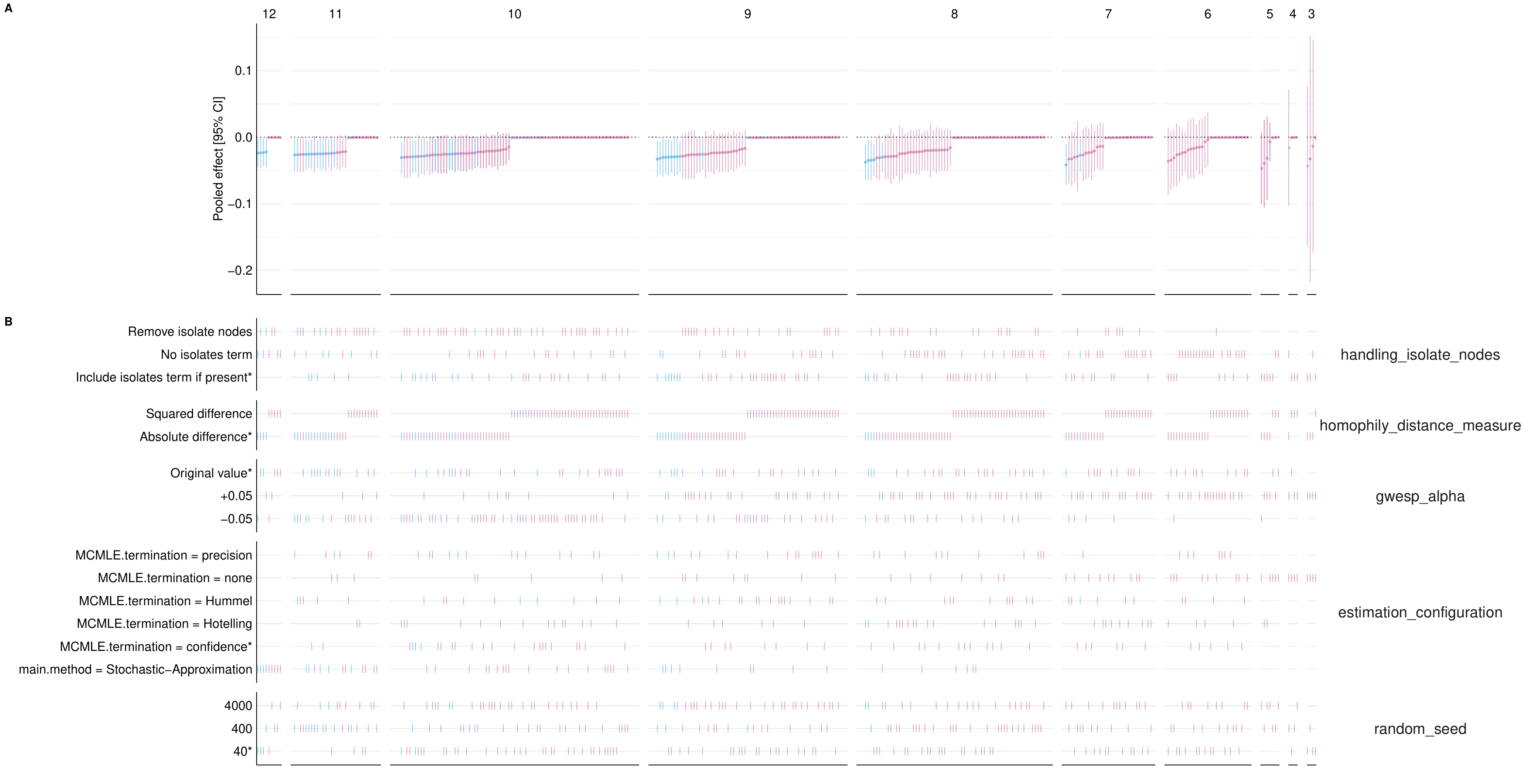}
    \caption{Results of the multiverse analysis of Study 2 in \cite{beffel2023understanding}.
        Panel~A displays the estimates (pooled effect sizes) of the prosocial score homophily distance from the meta-analysis as dots and the corresponding $95\%$ CIs as vertical bars.
        The columns display how many ERGMs of the 34 classroom networks achieved acceptable convergence and goodness-of-fit.
        Panel~B shows the specific decision combinations that were used for a specific effect shown in Panel~A. Blue represents decision combinations for which the $95\%$ CI is negative (same as the original direction); pink represents decision combinations for which the $95\%$ CI overlaps with $0$.}
    \label{fig_beffel_multiverse_extended_psh}
\end{sidewaysfigure}

In the original study, the meta-analyses were done on the 12 out of 34 classroom ERGMs that had acceptable convergence and goodness-of-fit to the data.
In our multiverse analysis, we investigate if certain decision combinations lead to the inclusion of more or less classroom data.
The results are shown in Figures~\ref{fig_beffel_multiverse_extended_ps} and~\ref{fig_beffel_multiverse_extended_psh}, where the specification curves are partitioned according to the number of classroom ERGMs included in the meta-analysis. As shown at the top of Figures~\ref{fig_beffel_multiverse_extended_ps} and~\ref{fig_beffel_multiverse_extended_psh}, the number of classroom ERGMs varies between 3 and 12. Among all decision combinations, no additional classroom data is included.
The number of included ERGMs expectedly have direct impact on the variance of the pooled effect sizes estimates of the meta-analyses, since less data and fewer ERGMs increase uncertainty. This is reflected by the increasingly larger CIs as the number of networks decreases in Figures~\ref{fig_beffel_multiverse_extended_ps} and~\ref{fig_beffel_multiverse_extended_psh}.
For Study 1, all CI overlapping are concentrated at the two smallest numbers of included ERGMs decreases (3 and 4).
For Study 2, the fraction of CI overlapping increases as the number of included ERGMs decreases.

\subsection{Challenges with ERGM}

Several challenges emerged when we conducted the ERGM multiverse analysis of \cite{beffel2023understanding}.

Multiverse analysis calls for better data quality as its core purpose is to examine how robust results are across plausible decision combinations.
In the case of \cite{beffel2023understanding}, the network data and prosocial scores were aggregated from the original survey data, and this aggregation itself involved multiple analytical decisions. However, since the raw survey data is not publicly available, we could not systematically vary and assess those upstream decisions as part of the multiverse analysis.
In addition, the ``prosocial score'' as a node attribute contained missing values in three classrooms. As a result, fitting ERGMs to these classroom networks always returned with errors.

Given network data, an ERGM is not guaranteed to converge out-of-the-box~\citep{hunter2008ergm}.
Even if an ERGM converges, it is not guaranteed to have good fit to the data~\citep{hunter2008goodness}. In practice, researchers often rely on adjusting model parameters and settings to help achieving model convergence~\citep{introduction_ergm, jones_ready_hazel_ergm_intro}.
It is also recommended to run the models with multiple chains or multiple times to ensure consistent results, and to set the random seed in the code for reproduction and replication of results in publications~\citep{jones_ready_hazel_ergm_intro}.

ERGMs with geometrically-weighted terms such as GWESP can be difficult to converge. \cite[][``Edgewise shared partnerships, or triangles revisited'' section]{jones_ready_hazel_ergm_intro} suggest \enquote{running models with different values of the decay parameter and choosing the decay parameter that gives the best model fit} a common work-around for this issue.

We argue that while the entire process of tuning parameters and settings to achieve model convergence and good fit to the data is not wrong and often necessary, it typically involves extensive trials and errors. These steps and their immediate results are rarely documented and reported systematically, which makes it difficult to evaluate how much uncertainty is introduced.

Multiverse analysis, instead, embraces this uncertainty by exposing the trials and error steps and making them transparent. Instead of reporting only a single, successfully converged specification, a multiverse documents where models converge or fail, how fit and effect estimates vary across specifications, and which conclusions remain robust. This makes the dependence of results on convergence-tuning decisions explicit and allows readers to assess how much of the substantive claim survives once this often-overlooked layer of uncertainty is taken into account.

\section{Case Study 3: \cite{Birkenmaier2025}}\label{case_study_3}

\subsection{The Original Study}

\cite{Birkenmaier2025} propose a computational framework to infer politicians' personality traits---agency and communion---in German speaking public communication. Using parliamentary speeches, social media posts, and broadcast interviews, they compare automated means for measuring said traits. 
Their data and code are publicly available, enabling direct computational replication.
They carefully evaluated automated approaches for measuring agency and communion in texts using a manually coded validation set. The authors compared three kinds of algorithms: 1) SVMs with a simple bag-of-words representation, 2) fine-tuned encoder-only transformer models, and 3) prompt based LLMs with few- and zero-shot configurations. The authors also tried using different training data compositions. They distinguish between explicit (or strong) and implicit (or weak) cues that signal the measured personality traits. With this notion they constructed two different sets of training and validation datasets, where the first variation contains only strong cues and the second contains both strong and weak cues. Furthermore, they also tried training with balanced and unbalanced data, where the unbalanced reflects the actual distribution of labels found in their corpus. The authors concluded that prompting Deepseek-V3 in a few-shot scenario yielded the highest $F_1$ score and selected this model for their subsequent analysis. Although they have trained multiple ML models, the authors' justification for the choice of only using Deepseek-V3 was that it was the ``best-performing model that successfully passed all previous validation steps.'' \citep[p.~12]{Birkenmaier2025}.

Based on these results, the authors measure the two personality traits in 15,000 randomly sampled sentences. In a subsequent analysis, \cite{Birkenmaier2025} investigate the alignment of the measured personality traits with political ideologies. They use the Chapel Hill Expert Survey (CHES) which maps all mainstream political parties in Germany according to two dimensions that comprise the left-right continuum: the economic left–right and the cultural libertarian–authoritarian dimension \citep{jolly:2022:chapel}. In their approach, they calculate the share of each personality trait at the sentence level for every politician. Next, they aggregate the mean for each party and quantify the measurement uncertainty with the standard deviation. Their findings suggest that \enquote{signaling of communion is systematically more prevalent among politicians from economically and culturally left-leaning parties} \citep[p.~12]{Birkenmaier2025}, as indicated by a regression line with a negative slope; or a negative regression coefficient for the libertarian–authoritarian dimension of a party in predicting the share of communion content. They also note that the agency dimension remains inconclusive and cannot be directly mapped to ideological positioning. 

\subsection{Multiverse Analysis}

We identified several explicit and implicit decisions in the original study. \cite{Birkenmaier2025} justified the choice of algorithm and training data compositions based on their attempts with various settings. However, for their ML attempts they only selected one pre-trained encoder-only transformer model, had limited variations on SVM settings, and relied on a narrow choice of LLMs. Therefore, we systematically vary the preprocessing steps and settings for SVMs, add two more encoder-only transformers \citep{chan2020germanslanguagemodel,wunderle2025gbert}, and also investigate the performance of locally hosted LLMs (GPT-OSS, Teuken, LLaMA3). A summary of decisions and variations can be found in Table~\ref{tab:birkenmaier_sml_choices}. Unlike in the original study where only the ML algorithm with the highest $F_1$ score (i.e. Deepseek-V3) was used, we conduct the subsequent statistical analysis with all possible ML algorithm choices.

\begin{table}
\centering
\caption{Overview of the decision combinations for the multiverse analysis of~\cite{Birkenmaier2025}: analytical decision considered and their defensible options.} 
\label{tab:birkenmaier_sml_choices}
\begin{tabularx}{\textwidth}{p{5cm} X}
\toprule
Analytical decision & Defensible options\\ \midrule
Training / test data composition$^a$ & (1) Balanced\newline
                                   (2) Unbalanced\\
Verbal cues & (1) Weak \& strong \newline
              (2) Strong\\
ML algorithm & (1) LLM prompting \newline
            \textit{LLM}$^b$\newline
            (i) Gemma3-27b\newline
            (ii) GPT-OSS-120b\newline
            (iii) Teuken\newline
            (iv) LLaMA3-8b\newline
            \textit{System Prompt}\newline
            (i) None\newline
            (ii) Expert coder\newline
            \textit{Prompt variation}\newline
            (i) Few-shot without the ``none'' category\newline
            (ii) Few-shot with the ``none'' category\newline
            (iii) Zero-shot\newline\newline
            (2) SVMs \newline
            \textit{Casing} \newline
            (i) Lowercasing\newline
            (ii) No lowercasing \newline
            \textit{Stemming}\newline
            (i) No stemming\newline
            (ii) Stemming\newline
            \textit{Stopwords}\newline
            (i) Removing stopwords\newline
            (ii) Keeping stopwords\newline
            \textit{Classifier outcome}\newline
            (i) Best classifier only\newline
            (ii) Ensemble\newline\newline       
            (3) Fine-tuned encoder\newline
            \textit{Pretrained model}\newline
            (i) XLM-RoBERTa\newline
            (ii) GermanBERT\newline
            (iii) ModernGBERT\newline
            \textit{Training epochs}\newline
            (i) $5$\newline
            (ii) $3$ \newline
            \textit{Learning rate}\newline
            (i) $0.00002$\newline
            (ii) $0.00005$\\
Corpus sampling & (1) $15000$ random \newline 
                  (2) $5000$ random \newline
                  (3) Stratified \\
Statistical testing & (1) Regression model (aggregated at the party level) \newline
                      (2) Logistic mixed-effects model\\ 
\midrule
Total specifications & $2 \times 2 \times (2 \times 2 \times 2 \times 2 + 3 \times 2 \times 2) \times 3 \times 2 + 2 \times (4\times2\times3)\times2\times3 = 960$ \\ \bottomrule
\end{tabularx}
\vspace{5pt} 
    \footnotesize
    \textit{Note:} Defensible option (1) for each analytical decision is the original settings.\newline
    $^a$ LLMs do not need to consider this decision.\newline
    $^b$ The original study employed additional proprietary, non-locally hosted LLMs (Deepseek-V3 and GPT-4o). We did not re-run them because they are fundamentally not computationally reproducible \citep{schoch2024computational}.
\end{table}

The next implicit decision is the sampling procedure from the studied corpus. The authors collected a corpus comprised of 44,905 sentences, of which they took a random sample of 15,000 sentences. We argue that other sampling strategies are equally plausible, thus, we also take a smaller sample of 5,000 sentences as well as a stratified sample, where we take 300 sentences per political party ($7 \times 300 = 2,100$). 

Finally, instead of merely aggregating the measurements first at the person level and then at the party level, we introduce another statistical analysis with a logistic mixed-effects model which can account for the hierarchical structure of the data.

All decisions together yield 960 decision combinations (see Table \ref{tab:birkenmaier_sml_choices}).

\subsection{Results}

\begin{sidewaysfigure}
    \centering
    \includegraphics[width=\linewidth]{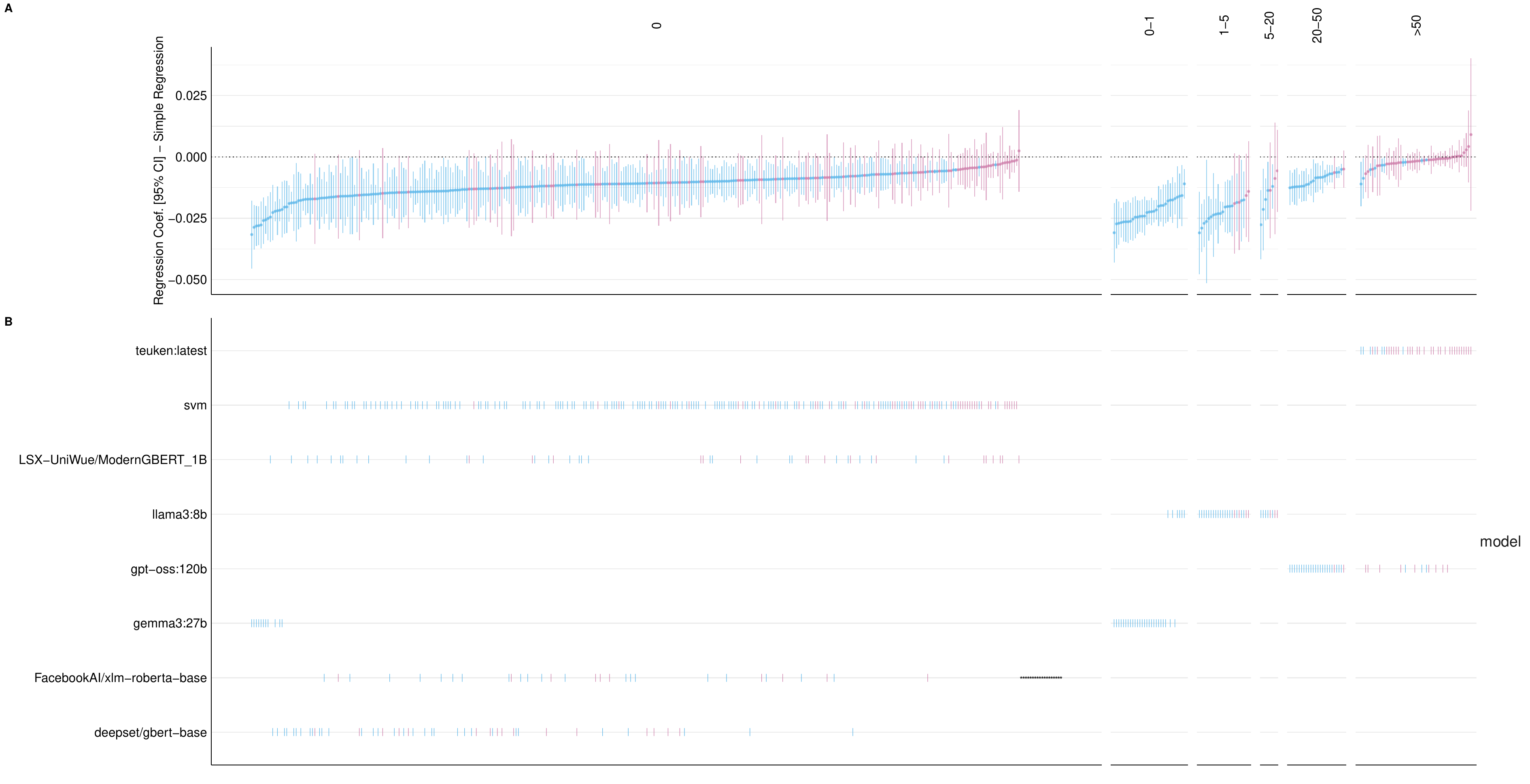}
    \caption{Results of the multiverse analysis of \cite{Birkenmaier2025}: Linear regression.
        Panel~A displays the estimates (regression coefficients of the libertarian–authoritarian dimension) as dots and the corresponding $95\%$ CIs as vertical bars.
        Panel~B shows the ML algorithms that were used for a specific effect shown in Panel~A. Blue represents decision combinations for which the $95\%$ CI is negative (same as the original direction); pink represents decision combinations for which the $95\%$ CI overlaps with $0$.}
    \label{birkenmaier_regression}
\end{sidewaysfigure}

\begin{sidewaysfigure}
    \centering
    \includegraphics[width=\linewidth]{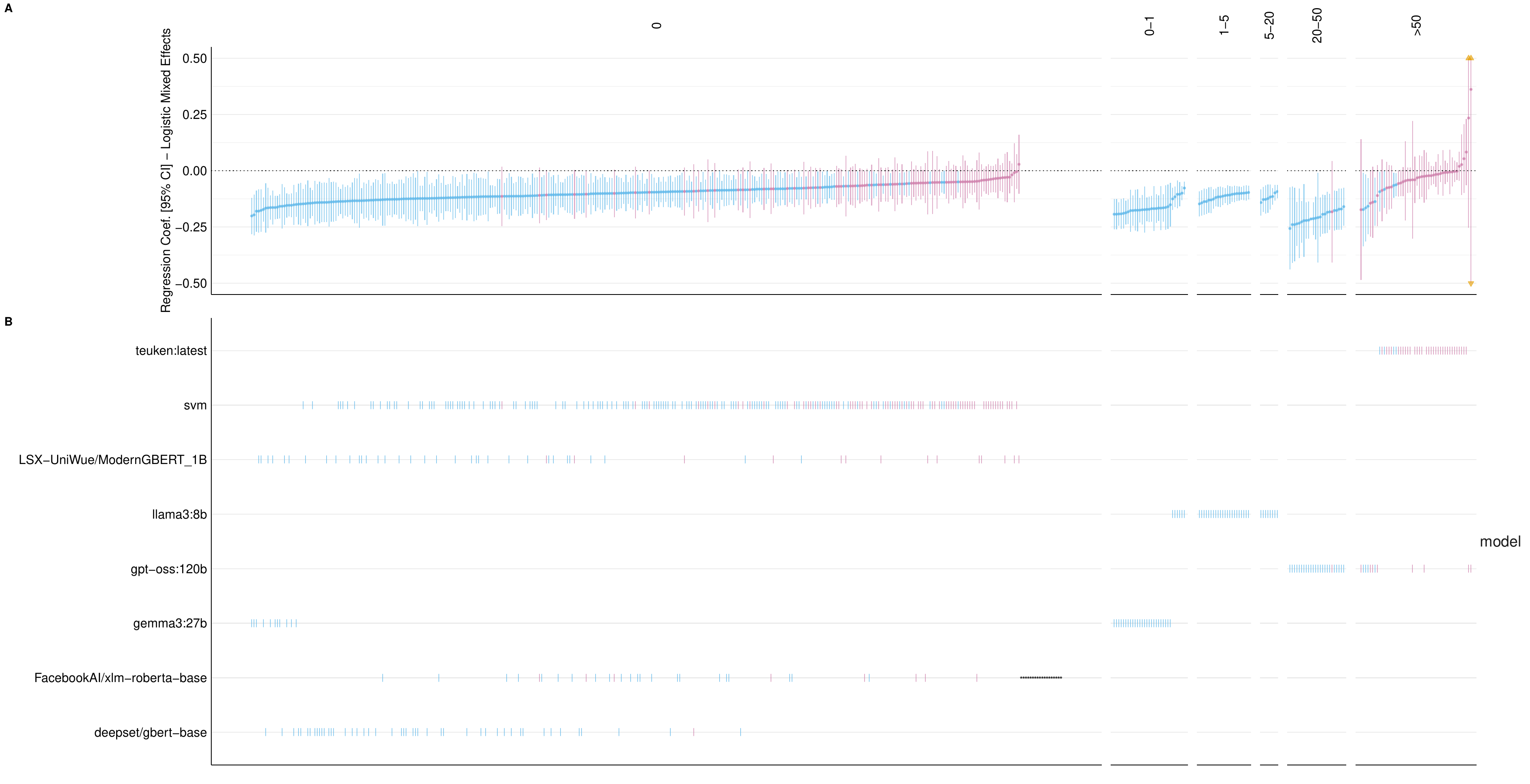}
    \caption{Results of the multiverse analysis of \cite{Birkenmaier2025}: Logistic mixed-effects regression.
        Panel~A displays the estimates (regression coefficients of the libertarian–authoritarian dimension) as dots and the corresponding $95\%$ CIs as vertical bars.
        Panel~B shows the ML algorithms that were used for a specific effect shown in Panel~A. Blue represents decision combinations for which the $95\%$ CI is negative (same as the original direction); pink represents decision combinations for which the $95\%$ CI overlaps with $0$.}
    \label{birkenmaier_mixed}
\end{sidewaysfigure}

As the linear regression and logistic mixed-effects regression have different estimands (see ``Statistical testing'' in Table \ref{tab:birkenmaier_sml_choices}), we follow \cite{DelgiudiceGangestad2021} and separate the two in the multiverse analysis (Figures~\ref{birkenmaier_regression} and~\ref{birkenmaier_mixed}). For clarity, we present the specification curves only displaying all estimates organized by ML algorithm, as other analytical decisions contribute relatively little to the observed variation. The complete specification curves are presented in Appendix~\ref{secA1}. Overall, the directionality of the effects in the multiverse analysis is largely in agreement with the findings of the original study.

While the majority of estimates is negative, we observe fewer estimates crossing zero in the logistic mixed-effects model (Figure~\ref{birkenmaier_mixed}). At the same time, there is notable variation depending on the sampling technique: in particular, stratified sampling leads to most estimates to cross zero. We also observe that the ML algorithm explains part of the variation in the estimates. One possible explanation is the failure rate, which is the proportion of text units that could not be classified at all (see section ``Challenges with Text Analysis'' below). Only SVM, the fine-tuned ModernGBERT, and GBERT produced predictions for every unit of text input. Therefore, we stratified the estimates by the failure rate in the two specification curves (Figures~\ref{birkenmaier_regression} and~\ref{birkenmaier_mixed}). In extreme cases, the algorithm did not provide any prediction for either the prompt and/or the input data, despite retrying. Because of that, the subsequent regression analysis failed (as indicated by the black stars in the specification curves). For some LLMs, the failure rate is unacceptably high; for example, Teuken with all cases had a higher than 50\% failure rate. The interval estimates are therefore unacceptably wide. One should deem these estimates as untrustworthy. Importantly, missing rates do not appear to be random. In all cases, we observed that the missing rates were not evenly distributed among all parties (which determine the independent variable: left-right placement).

Despite the failure rate, ML algorithms appear to explain the variations in estimates \textit{per se}. Gemma3 and LLaMA3, for instance, gave consistently much lower estimates than other ML algorithms.

\subsection{Challenges with Text Analysis}

Using multiverse analysis with preregistered decisions, our analysis, similar to Case 1 and Case 2, revealed that computational issues are much more common than one would infer from the literature. In particular, the fine-tuned XLM-RoBERTa failed completely in 36 cases due to suboptimal hyperparameter settings. While it is standard practice to tune hyperparameters and select the best-performing configuration based on a test dataset, such outright model failures are rarely documented in published studies. For several zero-shot LLMs, the algorithm likewise did not provide a usable prediction for either the prompt and/or the input text data, despite retrying. In some instances (e.g., Teuken, GPT-OSS, LLaMA3), the models did not adhere to the instructed output format, which specified a structured response to be parsed via regular expressions. Some of these failures can be attributed to deviations from the required format, but in other cases the models produced arbitrary text that could not be recovered even with adjustments to the parsing procedure.

In the two specification curves (Figures~\ref{birkenmaier_regression} and~\ref{birkenmaier_mixed}), we stratified the estimates by failure rate. Among the zero-shot LLMs, only Gemma3 consistently exhibited a low failure rate, remaining below 1\% across all specifications. At the same time, more traditional text analysis approaches such as SVM proved comparatively robust and performed on par with several deep learning–based methods.

Our overall finding is similar to \cite{baumann:2025:LLM} in that switching between LLMs can lead to different estimates, underscoring that these models cannot be treated as interchangeable in applied research. However, it remains difficult to justify \textit{ante factum} which model should be preferred, not least because new LLMs are released faster than they can be systematically validated and our results indicate that the size of the LLM does not correlate with better performance. Additional design choices may further affect performance. For example, although the original training data include preceding and following sentences, this contextual information was not used in the original study and thus not incorporated here; leveraging such context could plausibly improve transformer-based models, while potentially degrading performance for methods such as SVM. Similarly, we did not systematically vary prompting strategies for zero-shot and few-shot models. The differences between prompting templates for ``strong'' versus ``weak + strong'' signals are relatively minor, but a more thorough exploration lies beyond the scope of the present study.

Taken together, these results suggest that \cite{Birkenmaier2025} substantive conclusions could likely have been reached using more modest text analysis methods (i.e., SVM). At the same time, it is important to note that the authors do not advance strong substantive claims, and instead focus primarily on the measurement of personality traits (agency and communion), which represents a meaningful contribution in its own right.

\section{General Discussion}\label{general_discussion}

Across the three case studies, multiverse analysis proved useful in different but complementary ways. In Case Study 1, the main substantive finding remained robust across a range of alternative specifications, even though some combinations led to computational problems. In Case Study 2, robustness depended more clearly on the outcome under consideration: one hypothesis remained comparatively stable, whereas the other was sensitive to how key analytical choices were operationalized. In Case Study 3, the overall direction of the substantive pattern was broadly similar across many specifications, but the analysis also revealed substantial variation arising from model choice, failure rates, and downstream statistical decisions. Taken together, these cases show that the value of multiverse analysis in CSS lies not only in assessing whether a finding “holds,” but also in identifying which kinds of decisions matter most, where computational feasibility shapes the evidentiary base, and how uncertainty enters at different stages of the analytical pipeline. In this paper, we have demonstrated the utility of multiverse analysis for research in CSS through three case studies. Furthermore, we have shown how multiverse analysis can be applied across several research settings common in CSS and have discussed challenges and possible solutions. Throughout the manuscript, the advantages of multiverse analysis become evident. Nevertheless, we also seek to raise awareness of what multiverse is not, of how it should not be used, and what its limitations and negative aspects may be.

Even though we are convinced that multiverse analysis can contribute to transparency and more insightful results in CSS, we caution the reader  not to consider multiverse analysis to be an all-purpose solution that eliminates all uncertainty and resolves all flaws within the research. Instead, we think of multiverse analysis as a tool to understand rather than eliminate uncertainty and as a valuable contribution for addressing what \cite{Mcelreath2020} called issues of the \textit{small world} (in contrast to issues of the \textit{large world}), meaning that multiverse analysis offers nuance and consideration of modeling assumptions. As such, readers and authors of multiverse analyses should remind themselves that multiverse analysis is not a safeguard against reasonable criticism. This is important because a potential danger of multiverse analysis is that it can create an impression of invulnerability and finality. Even an elaborate multiverse still rests on subjective decisions within a small world and can therefore be challenged, as illustrated by a recent debate about two different views on multiverse analysis on the harmful effects of social media use on mental health \citep{TwengeHaidtJoiner_2020, TwengeHaidtLozano_2022, OrbenPrzybylski2019, OrbenPrzybylski2020, OrbenPrzybylski2019a}. 

Furthermore, researchers should take care not to push multiverse analysis to the extreme. One can legitimately question whether a decision combination that includes a set of covariates in addition to the actual variable of interest still corresponds to the same research question \citep[see][]{RohrerHullmanGelman2026}. For instance, the multiverse analysis by \cite{ganslmeier:2025:E} includes an impressive array of 3.6 billion decision combinations. However, \cite{auspurg:2025:R} criticizes \cite{ganslmeier:2025:E}'s multiverse analysis for overstating model uncertainties, as the multiverse analysis includes mostly inferior decision combinations. As commented by \cite[][p.~2]{auspurg:2025:R}, multiverse analyses seeking quantity over quality ``[drown] valid models in misspecified ones, needlessly eroding trust in science.'' While there exist researcher degrees of freedom and their potential influence on scientific findings, ``flooding the zone'' with billions of questionable specifications is clearly not a solution. Such a strategy is also wasteful in terms of time, financial resources, and environmental costs.

Choosing appropriate decision variations is challenging and ultimately not merely a methodological endeavor. Instead, researchers with substantive knowledge in the area of investigation should drive the selection of decisions and decision combinations. Ideally, the entire multiverse will be checked by independent experts before conducting the corresponding analyses. We advocate that a registered report is a suitable option for this. Not only can experts judge the plausibility of decision combinations in the multiverse, they can also reject options and thereby prune the multiverse tree. After all, nonsensical or redundant decision combinations should not be included in the multiverse, both to keep the analysis substantively meaningful and to conserve time, resources, and environmental impact. Current CSS research employs increasingly resource-hungry methods (e.g., LLMs). As such, researchers have an ethical responsibility to manage the size of the multiverse and keep the study feasible. We will come back to the selection of decisions and decision combinations in Section \ref{sec_selection}.

\subsection{Epistemological and Normative Implications of Multiverse Analysis}

From an epistemological perspective, our three case studies suggest that multiverse analysis should not be understood as a procedure for eliminating uncertainty, but rather as a way of making one important layer of uncertainty visible, namely the dependence of results on defensible analytical choices \citep{SteegenTuerlinckxGelman_2016,DelgiudiceGangestad2021,RohrerHullmanGelman2026}. Across the three cases, the issue was not simply whether findings ``survived'' alternative specifications, but which kinds of specifications materially altered the substantive conclusion, and under what conditions \citep{simonsohn:2020:S,young2025multiverse}.
In this sense, robustness is not equivalent to certainty. Rather, robust findings are those whose main substantive interpretation remains comparatively stable across a justified set of alternatives, whereas non-robust findings reveal where analytical discretion, computational constraints, or operational choices exert decisive influence. Importantly, even a carefully designed multiverse cannot exhaust all relevant uncertainty, because the construction of the multiverse itself depends on prior judgments about what counts as a plausible, comparable, and substantively meaningful specification \citep{DelgiudiceGangestad2021,RohrerHullmanGelman2026}.

This has broader theoretical implications for CSS, where analytical decisions are often distributed across multiple layers of the research process, including data selection, operationalization, model specification, software defaults, benchmark conventions, and computational feasibility. As a result, the production of findings in CSS cannot be understood as a purely technical pipeline in which methods neutrally process data. Rather, methods, infrastructures, and defaults partly shape what the object of analysis becomes and which forms of evidence are rendered visible, credible, or negligible. The case studies examined here therefore show that multiverse analysis is not only a tool for sensitivity testing, but also a way of making the socio-technical conditions of knowledge production more explicit. This reading aligns with broader methodological work that treats CSS as inseparable from ethical, contextual, and infrastructural considerations \citep{radovanovic:2026:NCO}.

These observations also point to normative implications for research practice. A good multiverse analysis should not be equated with maximal scope or combinatorial exhaustiveness. Instead, it should be transparently bounded, theoretically justified, and explicit about both included and excluded analytical paths \citep{DelgiudiceGangestad2021,PipalSongBoomgaarden2023,RohrerHullmanGelman2026}. This is particularly important in CSS, where resource limitations, proprietary systems, and environmental costs place real constraints on what can realistically be varied in practice. Under such conditions, the goal is not to simulate every conceivable universe, but to justify a fair and substantively defensible subset of universes that is sufficient to probe the main vulnerabilities of a claim and to communicate both robustness and remaining uncertainty in a way that is honest to readers and to the wider research community.

Based on these normative expectations, the key to the constructive use of multiverse analysis ---for CSS and beyond--- lies in two practices: (1) Selection of reasonable alternative specifications; and (2) Fair communication of multiverse analysis results.

\subsubsection{Selection of reasonable alternative specifications}\label{sec_selection}

The most consequential decision about multiverse analysis is which specifications to include. As indicated by the critique by \cite{auspurg:2025:R}, including a large number of questionable specifications can incorrectly create an impression of non-robust results. At the other extreme, one might include only nearly identical specifications. As a thought experiment of calculating the mean of a dataset with a billion observations, one excludes one data point around the mean as a specification. This can be repeated a million times to create a million virtually identical specifications. The expected consistency can create a misleading impression of robust results, because more consequential specifications are not evaluated.

Beyond their statistical role, these decisions also structure which aspects of the phenomenon are foregrounded or backgrounded in the reported narrative, reinforcing the need to treat specification choices as substantive, not merely technical.

Multiverse analysts must strive to strike a balance between these two extremes. \cite{DelgiudiceGangestad2021} provide a framework for determining which analytical decisions to include. They distinguish between Type E decisions (principled equivalence) Type N decisions (principled nonequivalence), and Type U decisions (uncertainty). 

Type E decisions include alternatives that are comparable, e.g., alternative measures that have comparable validity. For instance, the selection of media outlets based on the cutoff points of 0 or 1000 articles is a Type E decision. Type E decisions are genuinely arbitrary and should be considered in the same multiverse. The adjective \textit{arbitrary} does not mean random, esoteric, or without justification as in the everyday language. Instead, it has the legal meaning of ``the decision could have been made differently based on individual discretion.''

Type N decisions, however, do not have equivalent alternatives but might be better in terms of theoretical or statistical justifications. These decisions usually lead to different estimands. For instance, the choice between regression model and mixed-effects model in Case Study 3 is a type N decision, because the mixed-effects models we proposed are clearly preferable for not discarding a lot of information and accounting for the natural hierarchical structure of the data. Usually, the different estimands introduced by type N decisions are not directly comparable. Type U decisions are similar to Type N, only one has insufficient information to determine which alternatives are better.

\cite{DelgiudiceGangestad2021} suggest that Type N and U decisions should be considered in separate analyses, an approach that we followed (Figures~\ref{birkenmaier_regression} and \ref{birkenmaier_mixed}). When considering Type N decisions, multiverse analysts should exclude less preferable alternatives, ``because they are expected a priori to yield deflated effects, biased effects, or estimates suffering from low power and/or precision.'' \citep[][p.~7]{DelgiudiceGangestad2021} This is, in our opinion, key to preventing the ``flooding the zone'' problem, as multiverse analysts must deliberate among themselves which decisions and their alternatives are equivalent or not; or using our parlance, whether they are defensible options. Every specification should be justified and criticizable. This would help prevent the combinatorial generation of billions of questionable specifications as pointed out by \cite{auspurg:2025:R}.

That said, the decision to classify the decisions is not as clear-cut. This ambiguity is more pronounced in CSS research. To illustrate the problem, we had a hard time determining whether the choices of ML algorithms in Case Study 3 are Type E or Type N. If ML algorithms were Type N, we should trim all ``less preferable'' alternatives \citep{DelgiudiceGangestad2021}. If we were to follow the benchmark logic in \cite{Birkenmaier2025}, i.e. one should select the ``best'' ML algorithm based on the $F_1$ score calculated from a test set, we should eliminate many algorithms that have shown by \cite{Birkenmaier2025} to be ``inferior'' in our multiverse analyses, e.g., SVM and encoders. If we would follow the logic of both \cite{Birkenmaier2025} and \cite{DelgiudiceGangestad2021}, our multiverse analyses should then only include alternatives that are ``better'' than their final choice of Deepseek-V3 algorithm. This benchmark logic is also how the mainstream thinking about the validity of ML-based measurements in CSS \citep[e.g.,][]{birkenmaier:2023:SSG,ziems-etal-2024-large,weber:2024:E}. \footnote{The same logic is also applied in NLP and industrial AI research.}
This is precisely why the classification of analytical decisions in CSS remains partly a matter of substantive and methodological judgment rather than benchmark performance alone.

Ultimately, we treat the choice among ML algorithms as type E. That also implies that the decision by \cite{Birkenmaier2025} to select \textit{only} Deepseek-V3 based on the $F_1$ metric calculated from the test set \textit{is} arbitrary. The modest edge of Deepseek-V3 based on $F_1$ over other ML algorithms notwithstanding, our justifications are twofold. First, we look at the ML procedure in the two-step procedure introduced in the Introduction. Although $F_1$ measures the accuracy according to a certain ground truth, the application-agonistic evaluation does not measure the biases of the ML-based measurement introduced in the second step, e.g., the distribution of misclassified cases \citep{fong:2020:MLP,teblunthuis:2024:MAC,baumann:2025:LLM}. The criterion of being ``better'' just by using $F_1$ is insufficient if one considers both the ML procedure and the subsequent regression analysis holistically. Second, at the time of \cite{Birkenmaier2025} there were already social science publications that measure the constructs of interest with other ML algorithms, e.g., with SVM \citep{ramey:2016:MEP}, which they cited. If we were at the same time point of \cite{Birkenmaier2025} conducting their study, we should then consider other ML algorithms as defensible by the hitherto existed social science literature. Our justifications could be objectionable. But the point precisely that such justifications must be made explicitly within a multiverse analysis as a mechanism to exclude questionable multiverse specifications. 

If one wanted to move in the other direction and include a more diverse set of reasonable specifications,  we propose a parallel framework that can be considered together with \cite{DelgiudiceGangestad2021}'s: by considering the function of a decision. In a prototypical (computational) social science study, analytical decisions are made for, broadly speaking, three purposes: (1) selecting what data to analyze (selectional decisions), (2) operationalizing a certain construct based on the raw data (operationalizational decisions), and (3) choosing the statistical method and its configuration  (statistical decisions). A good multiverse analysis should include a healthy mix of the three types.

Using Case Study 2 as an example, there are selectional decisions (``handling isolate nodes''), operationalizational decisions (``distance measure for homophily'', ``decay factor of GWESP $\alpha$'') and statistical decisions (``handling non-converging models'', ``random seeds''). Some of these decisions were not made explicitly by \cite{beffel2023understanding}, but software defaults, e.g., ``handling non-converging models.'' We found that these implicit decisions are quite common in (computational) social science research and there are also these decisions in the rest of the two case studies (``prior'' in Case study 1, ``learning rate'' in Case Study 3). It is worthwhile to be reflexive about these implicit decisions and examine whether they would influence the empirical findings. Taken together, these considerations suggest that the task is not only to decide which specifications are defensible, but also how to communicate their consequences in a way that does justice to both robustness and remaining uncertainty.

\subsubsection{Fair communication of multiverse analysis results}

Multiverse analysis is usually not a built-in feature of a primary study \citep[cf.][]{PipalSongBoomgaarden2023,ivanusch:2025:messages}. Like the three case studies included here, as well as publications such as \cite{ganslmeier:2025:E,simonsohn:2020:S}, multiverse analysis is usually applied as a post-publication analysis. Naturally, some post-publication multiverse analyses, especially those that find the primary studies to be non-robust to alternative specifications, are critical and could even declare the primary studies as untrustworthy. Multiverse analysis appears to have the final say. But is this a fair way to treat the original studies? From a communicative perspective, the question is not only whether a finding survives alternative specifications, but how multiverse results are framed so that readers can see both the limits of robustness and the forms of uncertainty that remain.

Unlike what the metaphor ``multiverse'' suggests, original study authors do not enjoy the second-mover advantage of post-publication multiverse analysts. Because science is rarely a linear process, primary study authors face greater uncertainty and may also need to proceed through trial and error in order to settle on the appropriate analytical paths. As we argue in the Introduction section, some decisions, such as circumventing computational issues, are practical challenges and should not be conflated with researcher degrees of freedom.

A fair way to think about multiverse analysis, especially in a post-publication settings, is as a structured tool for clarifying which defensible analytical choices materially affect substantive conclusions. The purpose of a multiverse analysis is not simply to declare an original finding ``wrong'' or ``invalid,'' but to identify the conditions under which it remains stable, becomes attenuated, or changes direction. Framed in this way, multiverse analysis is less accusatory and more useful for cumulative research, because it helps subsequent studies identify which analytical decisions are most consequential. In our case studies, for example, the number of dictionary matches in Case Study 1 and the choice of ML algorithm in Case Study 3 proved particularly consequential for the resulting empirical patterns.

In addition, we advocate that authors of primary studies should also conduct multiverse analysis either as a planned preregistered analysis or, at minimum, as a post hoc analysis. Incorporating multiverse analysis into the research process leads to a more reflexive analysis of the arbitrary choices across the analytical pipeline. In this sense, integrating multiverse analysis into primary studies is not only a methodological choice, but also a commitment to more transparent and fair communication of robustness and uncertainty.

\subsection{Conclusion}\label{conclusion}

In this article, we have demonstrated through three case studies, spanning Bayesian multilevel modeling, network generative modeling, and text-based ML, how multiverse analysis can be used in CSS to examine the robustness of empirical claims under a structured set of defensible analytical choices. Across these examples, the aim was not to construct an exhaustive universe of possibilities, but to identify when and how alternative specifications materially change substantive conclusions and where findings remain comparatively stable.

Taken together, our results suggest that the main value of multiverse analysis lies less in eliminating uncertainty than in making visible which layers of uncertainty arise from defensible analytical discretion, computational constraints, and operational choices. In this sense, robustness is not equivalent to certainty: robust findings are those whose main substantive interpretation remains comparatively stable across a justified set of alternatives. By contrast, non-robust findings indicate where further theoretical refinement, improved measurement, or different modeling strategies may be needed.

\backmatter

\bmhead{Acknowledgements}

We would like to thank \cite{beffel2023understanding,Birkenmaier2025,ChanRauchfleisch2023} for making their replication materials and data publicly available. This study would not be possible without their commitments to Open Science. 
We also would like to thank Arnim Bleier for the discussion during the conceptualization of this study.

We used the following R packages: arrow v. 21.0.0.1 \citep{arrow}, callr
v. 3.7.6 \citep{callr}, caret v. 7.0.1 \citep{caret}, caretEnsemble v.
4.0.1 \citep{caretEnsemble}, cowplot v. 1.2.0 \citep{cowplot}, doFuture
v. 1.2.1 \citep{doFuture_and_future}, dotenv v. 1.0.3 \citep{dotenv}, foreach v.
1.5.2 \citep{foreach}, furrr v. 0.3.1 \citep{furrr}, future v. 1.67.0
\citep{doFuture_and_future}, grid v. 4.5.3 \citep{cranr}, here v. 1.0.1 \citep{here},
lme4 v. 2.0.1 \citep{lme4}, metafor v. 4.8.0 \citep{metafor}, osfr v.
0.2.9 \citep{osfr}, parallel v. 4.5.3 \citep{cranr}, posterior v.
1.6.1
\citep{posterior2021b, posterior2022d, posterior2024c, posterior2024e, posterior2025a},
progressr v. 0.18.0 \citep{progressr}, quanteda v. 4.3.1
\citep{quanteda}, rio v. 1.2.4 \citep{rio}, rollama v. 0.2.2.9000
\citep{rollama}, rstan v. 2.32.7 \citep{rstan}, statnet v. 2019.6
\citep{statnet2008, statnet2018}, tidyselect v. 1.2.1
\citep{tidyselect}, tidyverse v. 2.0.0 \citep{tidyverse}, tinytest v.
1.4.3 \citep{tinytest}.

We modified and extended the code in the R package specr \citep{specr}.

We used the following Python packages: click \citep{click}, datasets \citep{lhoest-etal-2021-datasets}, pandas \citep{reback2020pandas}, polars \citep{ritchie_vink_2026_19699949}, sklearn \citep{scikit-learn}, torch \citep{paszke2019pytorch}, transformers \citep{wolf-etal-2020-transformers}.

\section*{Statements and Declarations}

\begin{itemize}
\item \underline{Funding:} Open Access funding will be enabled and organized by Projekt DEAL.
\item \underline{Competing interests:} All authors declare no competing interests.
\item \underline{Ethics approval and consent to participate:} Not applicable because we only used secondary data.
\item \underline{Consent for publication:} All authors consent to publish the manuscript.
\item \underline{Data availability:} We exclusively used secondary data. As such, we do not provide the data. However, our publicly available code entails scripts for automatically downloading these secondary data used in our analyses.
\item \underline{Materials availability:} All supplementary materials (e.g., preregistrations) are available at the following view-only OSF link:  \url{https://osf.io/qazst/overview?view_only=4b67107b464e413db71be73405f4c1fc}.
\item \underline{Code availability:} All code is publicly available at the following view-only OSF link: \url{https://osf.io/qazst/overview?view_only=4b67107b464e413db71be73405f4c1fc}.
\item \underline{Author contribution:} \textbf{Conceptualization:} M.L., J.S.; \textbf{Data curation:} M.L., J.S., P.B.; \textbf{Formal analysis:} M.L., J.S., P.B.; \textbf{Methodology:} M.L., J.S., P.B.; \textbf{Software:} M.L., J.S., P.B., C-h.C.; \textbf{Visualization:} M.L., J.S., P.B., C-h.C.; \textbf{Writing --- original draft:} M.L., J.S., P.B., C-h.C., D.R.; \textbf{Writing --- review \& editing:} M.L., J.S., P.B., C-h.C, D.R.
\end{itemize}

\bibliography{bibliography}

\begin{appendices}

\section{Complete specification curves of \cite{birkenmaier:2023:SSG}}\label{secA1}

The following specification curves present the complete multiverse analysis of \cite{birkenmaier:2023:SSG}. Because there are 960 decision combinations, we organized these specification curves first by ML algorithm and then by statistical testing. Please refer to Figures~\ref{birkenmaier_regression} and~\ref{birkenmaier_mixed} for the meanings of the two panels.

\begin{sidewaysfigure}
    \centering
    \includegraphics[width=\linewidth]{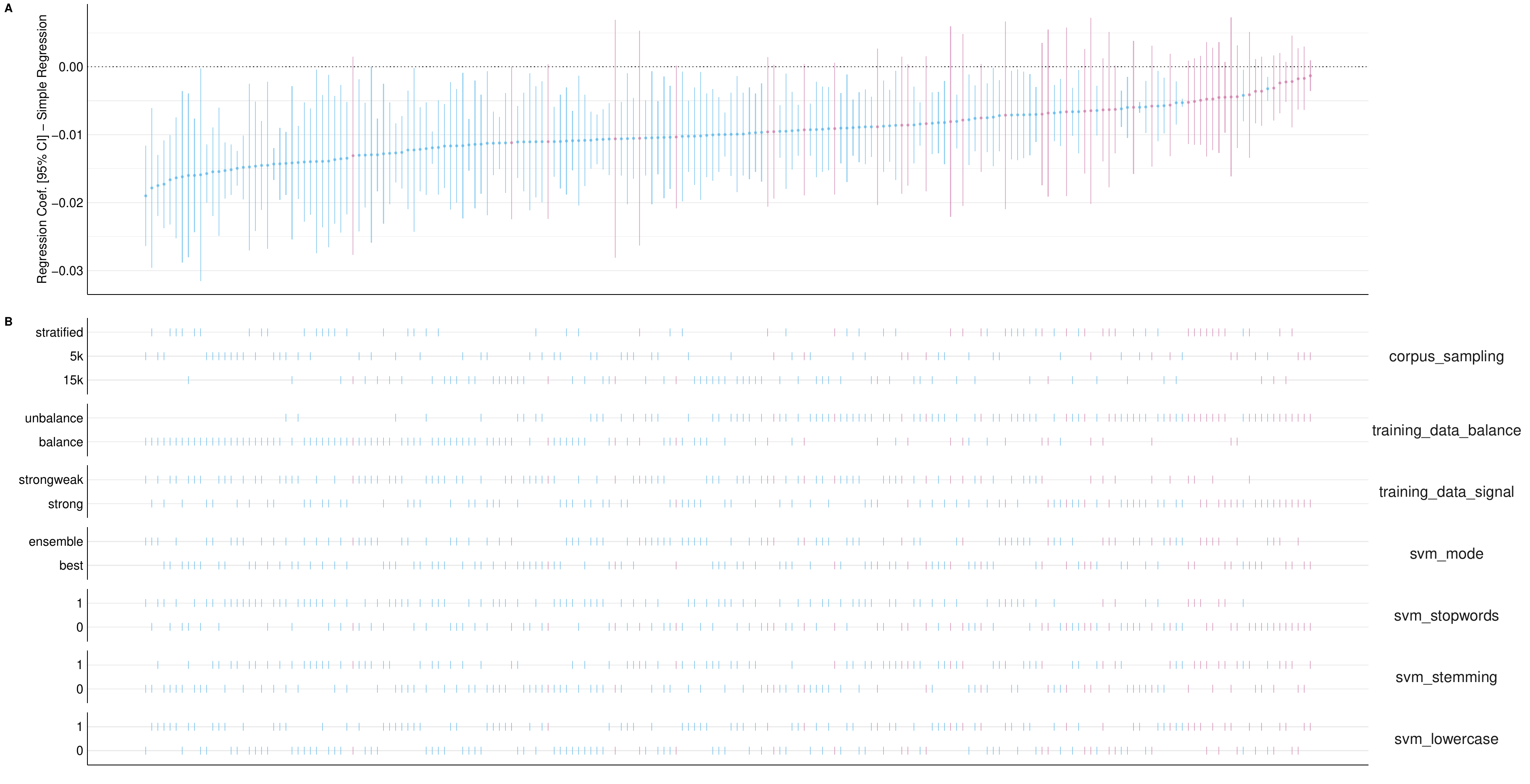}
    \caption{Results of the multiverse analysis of \cite{Birkenmaier2025}: SVM and linear regression.}
    \label{birkenmaier_regression_svm}
\end{sidewaysfigure}

\begin{sidewaysfigure}
    \centering
    \includegraphics[width=\linewidth]{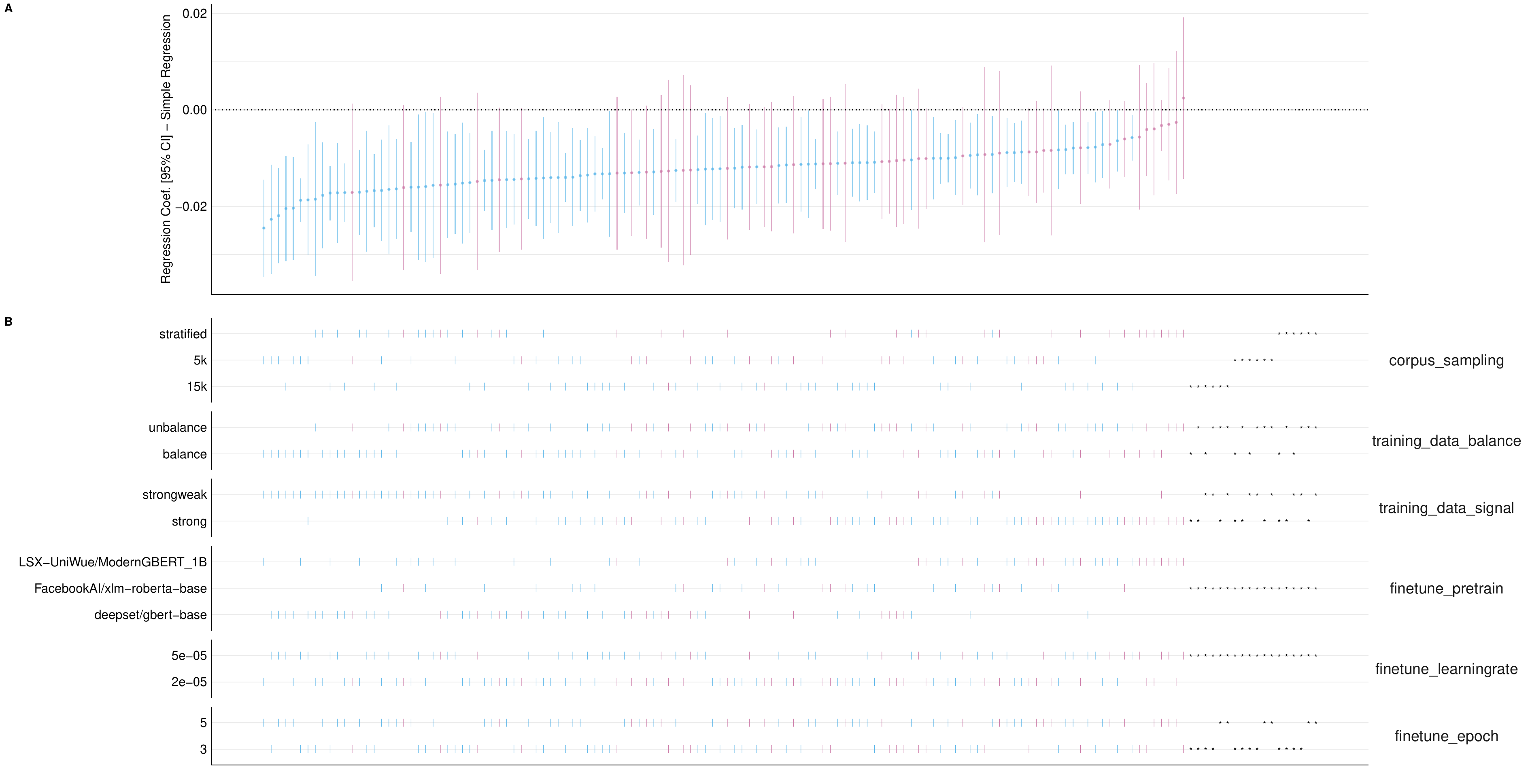}
    \caption{Results of the multiverse analysis of \cite{Birkenmaier2025}: Fine-tuned encoders and linear regression.}
    \label{birkenmaier_regression_finetune}
\end{sidewaysfigure}

\begin{sidewaysfigure}
    \centering
    \includegraphics[width=\linewidth]{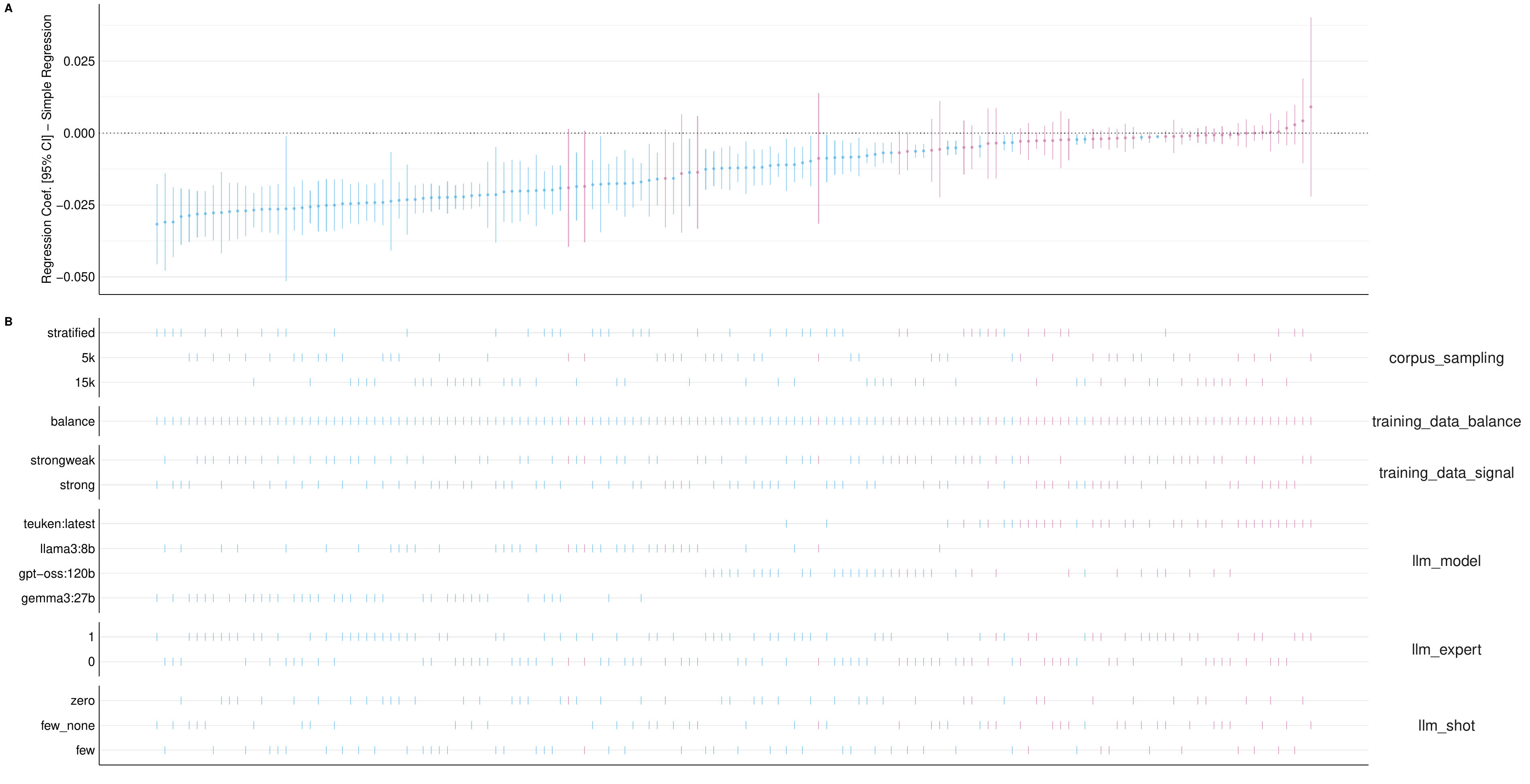}
    \caption{Results of the multiverse analysis of \cite{Birkenmaier2025}: LLMs and linear regression.}
    \label{birkenmaier_regression_llm}
\end{sidewaysfigure}

\begin{sidewaysfigure}
    \centering
    \includegraphics[width=\linewidth]{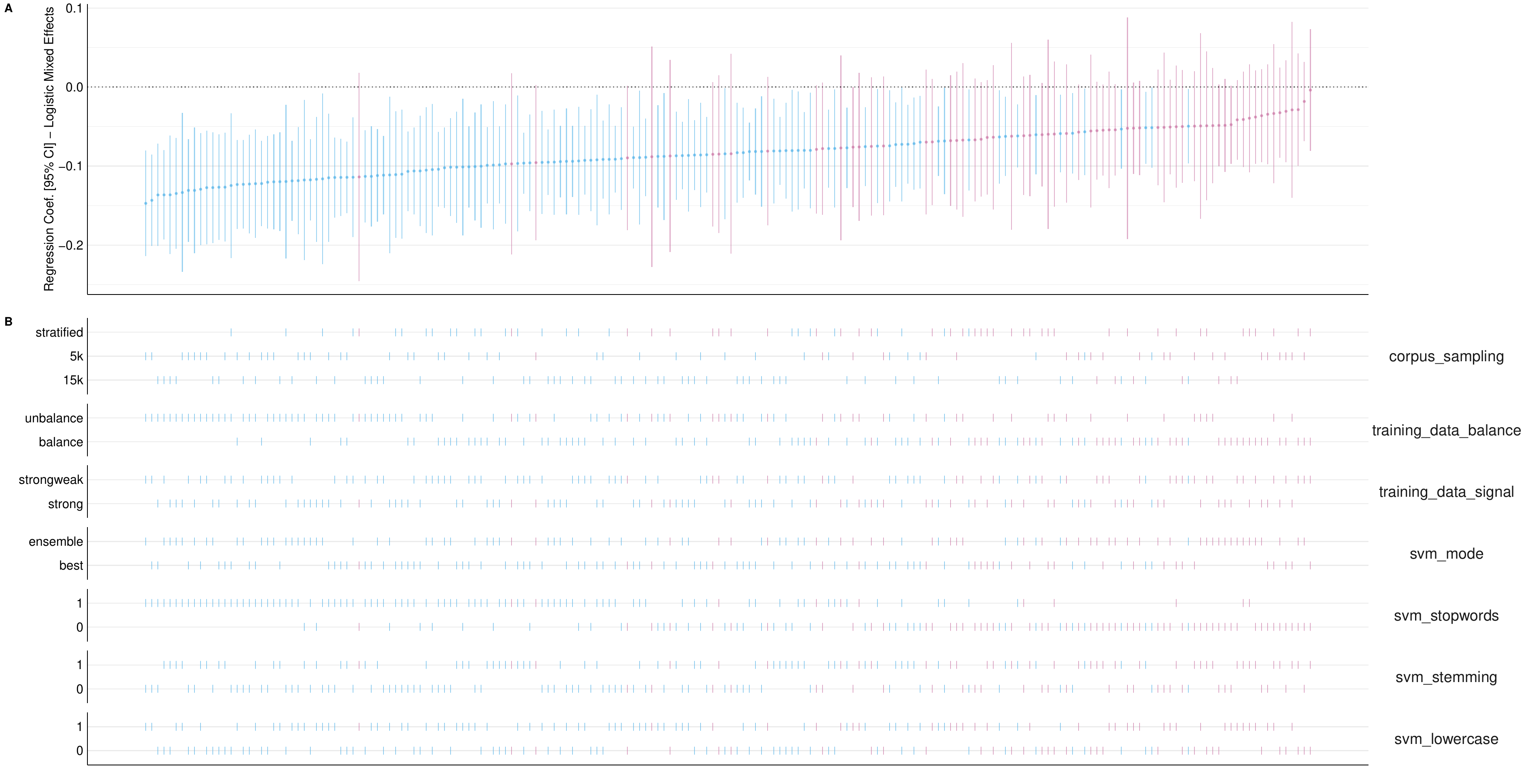}
    \caption{Results of the multiverse analysis of \cite{Birkenmaier2025}: SVM and logistic mixed-effects regression.}
    \label{birkenmaier_mixed_svm}
\end{sidewaysfigure}

\begin{sidewaysfigure}
    \centering
    \includegraphics[width=\linewidth]{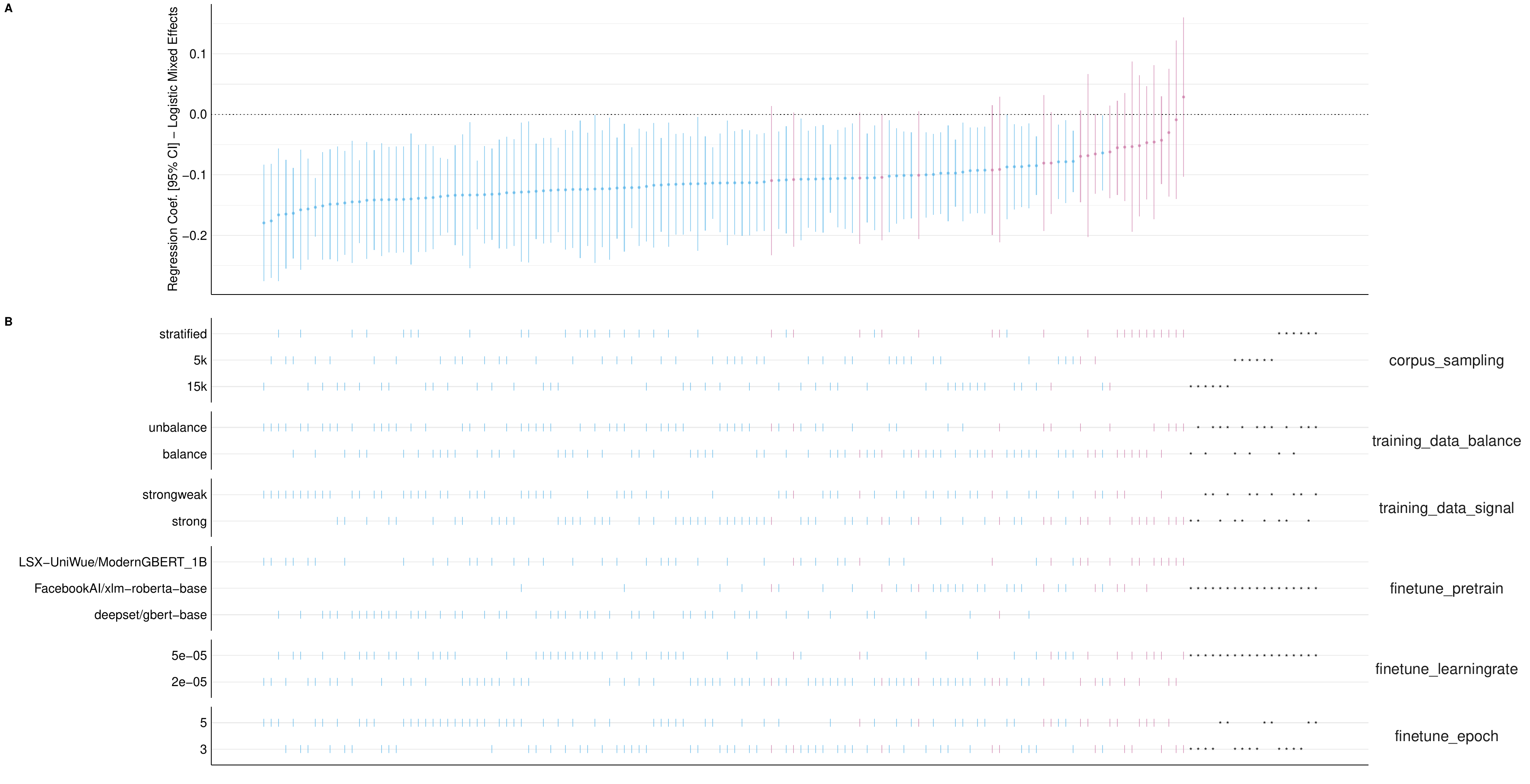}
    \caption{Results of the multiverse analysis of \cite{Birkenmaier2025}: Fine-tuned encoders and Logistic mixed-effects regression.}
    \label{birkenmaier_mixed_finetune}
\end{sidewaysfigure}

\begin{sidewaysfigure}
    \centering
    \includegraphics[width=\linewidth]{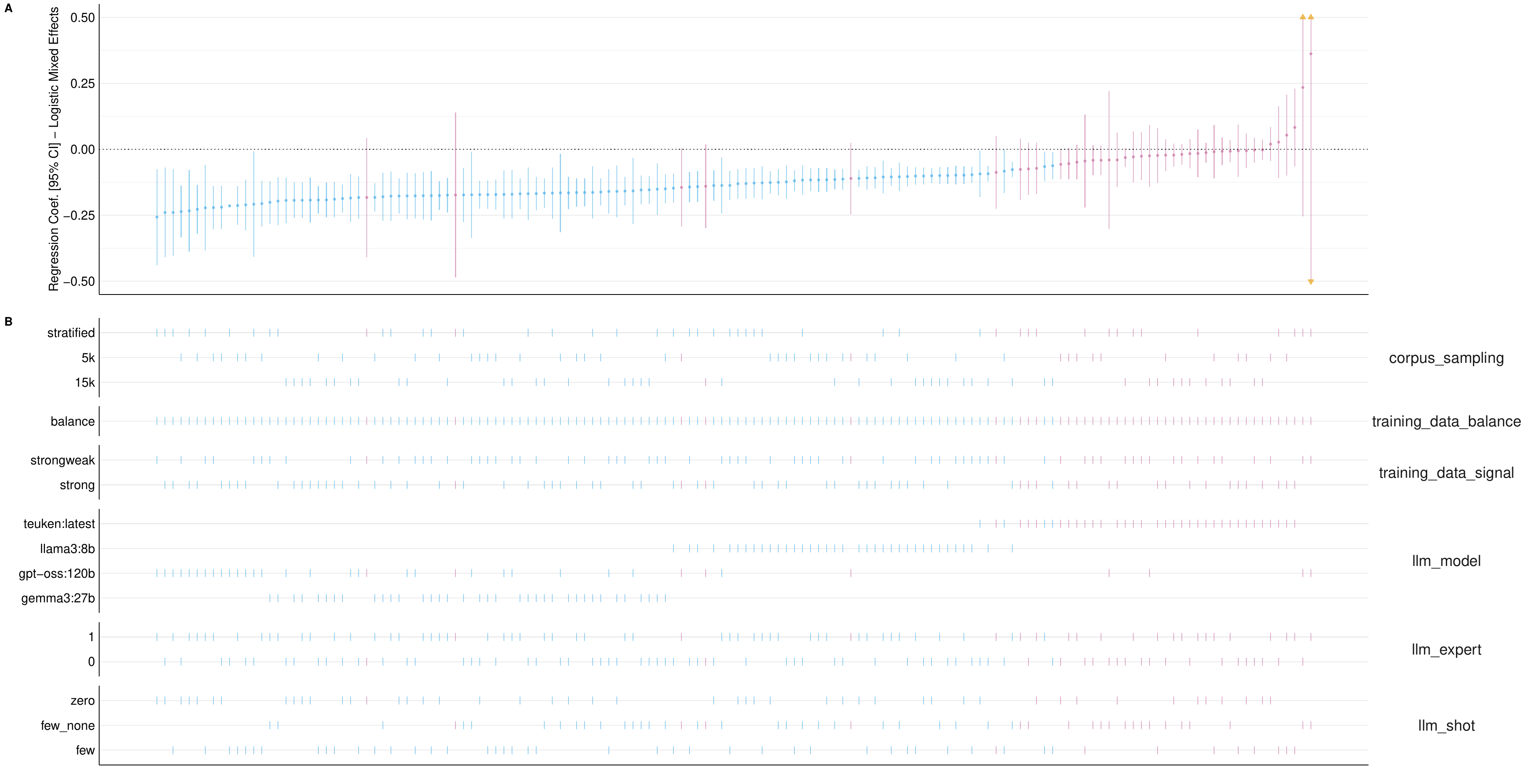}
    \caption{Results of the multiverse analysis of \cite{Birkenmaier2025}: LLMs and logistic mixed-effects regression.}
    \label{birkenmaier_mixed_llm}
\end{sidewaysfigure}

\end{appendices}

\end{document}